\documentclass[
]{aa}
\usepackage[varg]{txfonts}

\usepackage[dvipsnames]{xcolor}
\usepackage{amsmath}

\usepackage{balance} 

\definecolor{xlinkcolor}{cmyk}{1,1,0,0}
\usepackage[unicode, 
bookmarks=true,         
pdfnewwindow=true,      
colorlinks=true,    
linkcolor=xlinkcolor,     
citecolor=xlinkcolor,     
filecolor=xlinkcolor,  
urlcolor=xlinkcolor,      
final=true,
]{hyperref}

\hypersetup{
	pdftitle={Unravelling stellar populations in the Andromeda Galaxy},
	pdfauthor={Grzegorz Gajda et al.},
}

\makeatletter
\renewcommand*\aa@pageof{, page \thepage{} of \pageref*{LastPage}}
\makeatother

\bibpunct{(}{)}{;}{a}{}{,} 

\newcommand{\orcid}[1]{\href{https://orcid.org/#1}{\raisebox{3pt}{\protect\includegraphics[width=7pt]{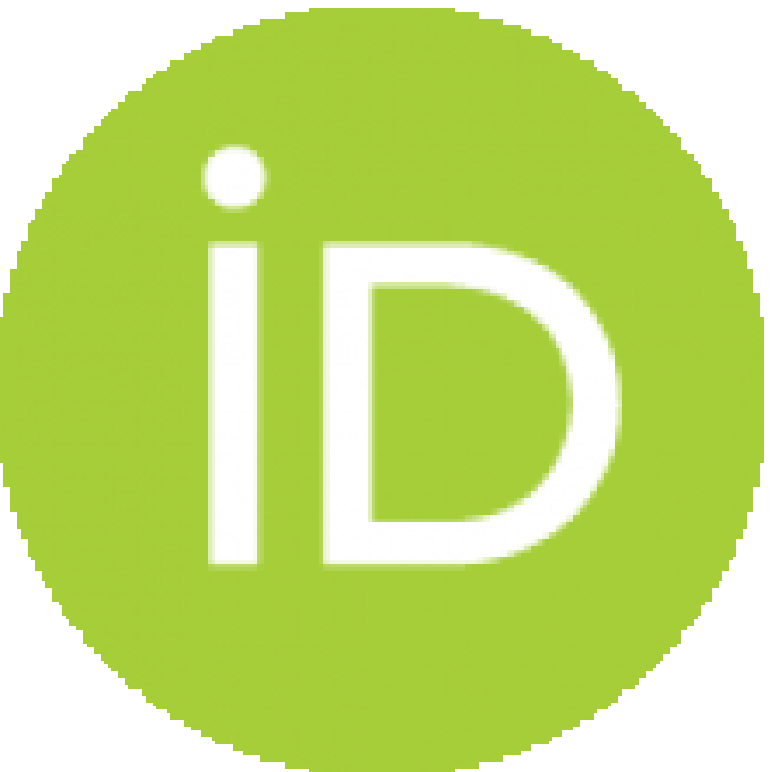}}}}

\begin{document}
	
	\title{Unravelling stellar populations in the Andromeda Galaxy}
	
	\titlerunning{Unravelling stellar populations in the Andromeda Galaxy}
	\authorrunning{Gajda et al.}
	
	\author{
		Grzegorz Gajda\inst{\ref{inst_mpe}}\thanks{\email{ggajda@mpe.mpg.de}}
		\orcid{0000-0002-9290-0473}
		\and
		Ortwin Gerhard\inst{\ref{inst_mpe}}
		\orcid{0000-0003-3333-0033}
		\and
		Mat{\'\i}as Bla{\~n}a\inst{\ref{inst_mpe}, \ref{inst_usm}}
		\orcid{0000-0003-2139-0944}
		\and
		Ling Zhu\inst{\ref{inst_shao}}
		\orcid{0000-0002-8005-0870}
		\and
		Juntai Shen\inst{\ref{inst_jtu},\ref{inst_shao}}
		\orcid{0000-0001-5604-1643}
		\and
		\\Roberto P. Saglia\inst{\ref{inst_mpe},\ref{inst_usm}}
		\orcid{0000-0003-0378-7032}
		\and
		Ralf~Bender\inst{\ref{inst_mpe},\ref{inst_usm}}
		\orcid{0000-0001-7179-0626}
	}
	
	\institute{
		Max-Planck-Institut f\"ur Extraterrestrische Physik, Giessenbachstrasse, D-85748 Garching, Germany\label{inst_mpe}
		\and
		Shanghai Astronomical Observatory, Chinese Academy of Sciences, 80 Nandan Road, Shanghai 200030, China\label{inst_shao}
		\and
		Department of Astronomy, School of Physics and Astronomy, Shanghai Jiao Tong University, 800 Dongchuan Road,\\Shanghai 200240, China\label{inst_jtu}
		\and
		Universit\"ats-Sternwarte M\"unchen, Scheinerstrasse 1, D-81679 M\"unchen, Germany\label{inst_usm}
	}
	
	\date{Submitted to A\&A 8 May 2020 / Resubmitted 15 December 2020 / Accepted 16 January 2021}
	
	\abstract{
		To understand the history and formation mechanisms of galaxies it is crucial to determine their current multidimensional structure.
		Here we focus on stellar population properties, such as metallicity and [$\alpha$/Fe] enhancement.
		We devise a new technique to recover the distribution of these parameters using spatially resolved, line-of-sight averaged data.
		Our chemodynamical method is based on the made-to-measure (M2M) framework and results in an $N$-body model for the abundance distribution.
		We test our method on a~mock data set and find that the radial and azimuthal profiles are well-recovered, however only the overall shape of the vertical profile matches the true profile.
		We apply our procedure to spatially resolved maps of mean [Z/H] and [$\alpha$/Fe] for the Andromeda Galaxy, using an earlier barred dynamical model of M31.
		We find that the metallicity is enhanced along the bar, with possible maxima at the ansae.
		In the edge-on view the [Z/H] distribution has an X shape due to the boxy/peanut bulge; the average vertical metallicity gradient is equal to $-0.133\pm0.006$ dex/kpc.
		We identify a metallicity-enhanced ring around the bar, which also has relatively lower [$\alpha$/Fe].
		The highest [$\alpha$/Fe] is found in the centre, due to the classical bulge.
		Away from the centre, the $\alpha$-overabundance in the bar region increases with height, which could be an indication of a thick disc.
		We argue that the galaxy assembly resulted in a sharp peak of metallicity in the central few hundred parsecs and a more gentle negative gradient in the remaining disc, but no [$\alpha$/Fe] gradient.
		The formation of the bar lead to the re-arrangement of the [Z/H] distribution, causing a flat gradient along the bar.
		Subsequent star formation close to the bar ends may have produced the metallicity enhancements at the ansae and the [Z/H] enhanced lower-$\alpha$ ring.
	}
	
	
	\maketitle

	\section{Introduction}
	The central parts of disc galaxies are occupied by bulges, which can be classified into two broad categories \citep{kormendy_kennicutt2004, fisher_drory2016}.
	The classical bulges were probably formed very early on, from violent early gas-rich mergers or mergers within clumpy disks \citep{hopkins2009, brooks_christensen2016, bournaud2016}.
	Boxy-peanut and disky bulges are thought to be built through evolution of the disc component, triggered by bar formation \citep{kormendy2013, fragkoudi2020}.
	It has been found that different types of bulges can coexist in a single galaxy \citep[e.g.][]{erwin2015}.
	In such a case, the bar will transfer some of its angular momentum and spin-up the classical bulge \citep{saha2012,saha2016}.
	
	A major channel of bar formation is a global instability.
	$N$-body simulations showed early-on that disc galaxies are prone to development of elongated structures in their centres \citep{miller1970, hohl1971}.
	Shortly after their formation bars thicken, acquiring a boxy/peanut (b/p) shape in the side-on view \citep{combes_sanders1981, combes1990}.
	This is a result of another instability called \emph{buckling} \citep{raha1991, athanassoula_misiriotis2002, debattista2006}.
	An alternative explanation of this process is thickening through a vertical resonance \citep{combes1990, quillen2014, sellwood_gerhard2020}.
	The vertically extended part of the bar constitutes the b/p bulge (\citealt{lutticke2000,athanassoula2005, erwin_debattista2013}, see also \citealt{athanassoula2016bulges} for a recent review).
	More about the theoretical understanding of bar physics can be found in the reviews by \citet{athanassoula2013review} and \citet{sellwood2014}.
	The fraction of barred galaxies grows with cosmic time, starting from $\sim\!10\%$ at $z=1$ \citep{sheth2008, melvin2014} to about $50$--$70\%$ in the local Universe \citep{skibba2012, erwin2018}.
	Also the abundance of the b/p bulges grows with time \citep{kruk2019}, reaching $\sim\!40\%$ at $z=0$ \citep[see also][]{lutticke2000}.
	
	The discussion about the impact of bars on stellar populations has not yet concluded.
	Some authors suggest that bars lead to higher metallicities in the galaxy centres \citep{moorthy_holtzman2006, perez_sanchez-blazquez2011}, while others do not identify significant differences \citep{jablonka2007, williams2012, cheung2015}.
	\citet{perez2009} found all types of metallicity gradients along bars: positive, flat and negative.
	\citet{williams2012} argue that gradients in bars are flatter than in the discs, suggesting that bar formation smears out pre-existing gradients.
	\citet{coelho_gadotti2011} conclude that bulges in barred galaxies are on average 4 Gyr younger than in unbarred ones.
	\citet{sanchez-blazquez2014} do not find any differences in the metallicity gradients in the outer parts of barred and unbarred galaxies, contrary to some simulation predictions \citep{friedli1994, minchev_famaey2010}.
	In the Milky Way, stars in the immediately surrounding disc appear to be slightly younger and more metal rich than in the bar region \citep[][]{hayden2015, bovy2019}.
	
	Our neighbour, the Andromeda Galaxy, is an excellent target for investigating the stellar populations in the centre of a~large galaxy.
	M31 has long been described as hosting a classical bulge \citep[e.g.][]{kormendy_bender1999, kormendy2010}.
	However already early-on, \citet{lindblad1956} posed that the central twist of the isophotes in M31 is caused by a bar.
	This argument was strengthened by \citet{athanassoula_beaton2006}, who compared barred $N$-body models to the near-infrared image of \citet{beaton2007} and concluded that the bar has a length of ${\approx}\,1300$\arcsec\ (${\sim}\,5$ kpc).
	\citet{blana2017} considered an array of models including both a classical bulge and a bar. 
	They concluded that the classical bulge contributes ${\sim}\,1/3$ of the mass of Andromeda's bulge, while the b/p bulge contribution is ${\sim}\,2/3$.
	\citet{blana2018} extended that work, modelling both the infrared image from \citet{barmby2006} and the kinematics derived by \citet{opitsch2018}, using the made-to-measure (M2M) technique.
	They concluded that the bar has length of ${\approx}\,4$~kpc and is oriented at $54.7\pm3.8\degr$ with respect to the line of nodes of M31.
	\citet{opitsch2018} provide a more extensive account of the evidence for the barred nature of M31.
	Several lines of evidence, such as the presence of the Giant Stream \citep[e.g.][]{sadoun2014,hammer2018} and several other substructures, the recent burst of star formation \citep{williams2015}, and the stellar age-velocity dispersion relation in the disc \citep[][]{bhattacharya2019} point to a recent ($\sim3$ Gyr ago) merger with a mass ratio approximately 1:5, which would likely also have left an impact on the distribution of the stellar populations in the inner regions of M31.
	
	Recently, a wide-field IFS survey of Andromeda was performed by \citet{opitsch2018}.
	Subsequently, \citet{saglia2018} analysed their spectra using Lick indices to derive stellar population properties for M31.
	They found that $80\%$ of their measurements indicated ages larger than $10$~Gyr.
	The metallicity along the bar was solar, with a peak of $0.35$ dex in the very centre.
	The [$\alpha$/Fe] enhancement was approximately $0.25$~dex \mbox{everywhere}, rising to $0.35$ dex in the centre.
	They proposed a~two-phase formation scenario, according to which at first the classical bulge formed in a quasi-monolithic way in parallel with the primeval disk. 
	Somewhat later, the bar formed and buckled into a b/p bulge while star formation continued not only in the disc, but also in the inner $2$ kpc.

	Galaxies are distant objects and we can observe them only in projection on the sky. 
	However, to really understand their structure we need to decipher the three-dimensional distribution of their components.
	It has been demonstrated that a \emph{deprojection} of the surface density is increasingly degenerate away from special cases such as a thin disk or an exactly edge-on axisymmetric system \citep{rybicki1987, gerhard_binney1996}.
	For triaxial systems already \citet{stark1977} illustrated the degeneracy by finding a sequence of ellipsoidal bulge models that would reproduce the observed twist between the bulge and the disk isophotes in M31, given a common principal plane.
	Besides the density distribution of the luminous and dark components, the distribution of the stellar population properties is also of interest.
	In particular, metallicity, elemental abundances and stellar ages are vital to understanding the evolution of galaxies.
	
	Here we consider the determination of the three-dimensional distribution of mean stellar population properties from the observational data for M31.
	We use the made-to-measure technique \citep{syer_tremaine1996} to incorporate the constraint that in dynamical equilibrium stellar population properties must be constant along orbits.
	In a standard M2M application one adjusts a particle model to a set of constraints by iteratively adjusting the masses of the particles. 
	The technique was adapted by \citet{delorenzi2007} to fit observational data through minimisation of a respective $\chi^2$ and implemented as the \textsc{nmagic} code.
	It has been used to study elliptical galaxies \citep{delorenzi2008, das2011}, the Milky Way \citep{portail2015m2m, portail2017a, portail2017b}, and Andromeda \citep{blana2018}.
	The M2M method was found by \citet{long_mao2012} to give similar results to the \citet{schwarzschild1979} modelling and to reproduce well analytically known distribution functions \citep{tagawa2016}.
	In particular, \citet{portail2017b} used distance resolved stellar parameter data to reconstruct the distribution of metallicities in the Milky Way bulge.
	The Schwarzschild orbit method has recently been extended to stellar population modelling as well \citep{poci2019, zhu2020}.

	We build on the M31 dynamical model by \citet{blana2018} and the stellar population analysis by \citet{saglia2018}.
	Our goal is to construct a three-dimensional model of metallicity and $\alpha$-enhancement in the Andromeda Galaxy.
	In Section \ref{sec_data} we present and discuss the available data based on Lick indices.
	Next, in Section \ref{sec_methods}, we present our newly developed technique, test it on a mock galaxy model and comment on the uncertainties introduced by the limited nature of the available information.
	Then we apply our method to M31, first to [Fe/H] in Section \ref{sec_met} and then to [$\alpha$/Fe] in Section \ref{sec_alpha}.
	We discuss our results and plausible origins of the observed trends in Section \ref{sec_discussion}, and finally summarise our conclusions in Section~\ref{sec_conclusions}.

	\section{Spatially resolved stellar population maps for M31}
	\label{sec_data}
	
	The Andromeda Galaxy is the closest large spiral galaxy, which is both an opportunity and a challenge.
	The close distance enables us to create very detailed maps of various quantities.
	On the other hand, M31 has a large size on the sky, thus one needs multiple visits or a survey with an extended sky coverage.
	
	As constraints for our model we use the publicly available stellar population properties derived by \citet{saglia2018} for the central regions of M31, based on the data collected by \citet{opitsch2018} with the VIRUS-W instrument \citep{fabricius2012virusw}.
	They covered the bulge area and sparsely sampled the adjacent disc along six directions.
	Spectra were rebinned to reach minimum $\mathrm{S/N}=30$ and the analysis yielded usable spectra for 6473 Voronoi cells.
	
	\citet{saglia2018} measured the absorption line strengths in the Lick/IDS system \citep{worthey1994}, using the following six indices: H$\beta$, Mg b, Fe5012, Fe5270, Fe5335 and Fe5406.
	To retrieve the stellar population parameters, they interpolated the models of \citet{thomas2011} on a finer grid, extending from $0.1$ to $15$ Gyr in age (in steps of $0.1$ Gyr), from $-2.25$ to $0.67$ dex in metallicity (in steps of $0.02$ dex) and from $-0.3$ to $0.5$ dex in $\alpha$-enhancement (in steps of $0.05$ dex).
	For each binned spectrum, \citet{saglia2018} compared the aforementioned indices to the grid of models and found that with the lowest $\chi^2$ value.
	The parameters of that model (age, [Z/H] and [$\alpha$/Fe]) were then assigned to this spectrum.
	Uncertainties were estimated by finding the range of models within $\Delta\chi^2\leq1$ with respect to the best-fit model.
	The errors of the metallicity and $\alpha$-abundance were floored at 0.01 dex.
	The reported mean uncertainties of [Z/H] and [$\alpha$/Fe] were, respectively, $0.04$ dex and $0.02$ dex.
	
	\citet{trager_somerville2009} found that metallicities derived from the Lick indices, so-called SSP-equivalents, follow the mass- or light-weighted metallicity of a given composite spectrum for model early-type galaxies. Metallicity obtained in this way slightly underestimates the true value by up to 0.1 dex and has scatter smaller than 0.1 dex.
	
	While it might not be surprising that one can treat the measured [Z/H] abundances as mass-weighted averages of the underlying stellar populations, certainly the [$\alpha$/Fe] abundance ratio needs more explanation.
	\citet{serra_trager2007} found that [$\alpha$/Fe] measured from the Lick indices well-reproduces a light-weighted mean of the stellar populations.
	\citet{pipino2006,pipino2008} argued that in the case of the $\alpha$-enhancement the light- and mass-weighted averages should give the same results because the distribution of [$\alpha$/Fe] should be relatively narrow and symmetric.
	Finally, [$\alpha$/Fe] is actually a logarithm of a ratio and can be transformed into a difference of two logarithms, namely [$\alpha$/H] and [Fe/H].
	Since [Fe/H] can be treated as mass-weighted, we suppose that one can extrapolate this to treat [$\alpha$/H] as mass-weighted too.
	Hence, we will treat [$\alpha$/Fe] as mass-weighted over the stellar populations along the line of sight.
	
	\citet{saglia2018} also derived the distribution of the stellar ages (SSP-equivalent) in the M31 bulge area.
	The map shown in their Fig.\ 13 is mostly featureless, especially in the (b/p) bulge region, i.e. it does not reveal any new structures.
	In particular, \citet{saglia2018} inspected simulations of a mix of stellar populations, based on the results of \citealt{dong2018}, and concluded that the bulge area is uniformly composed of a majority of old ($\geq 8$ Gyr) stars and a minority of younger ($\leq 4$~Gyr) stars. 
	Furthermore, the distribution of age differences is likely not well-resolved, because $\approx 40\%$ of the age measurements fall on the edge of the grid of models at $15$~Gyr. Therefore, we do not model the age distribution here. In general, modelling SSP-equivalent ages would be significantly more complex than what we aim for here, because these 
	SSP-equivalent ages are known to underestimate (with large scatter) light-weighted or mass-weighted ages when (relatively) younger components are present \citep{serra_trager2007, trager_somerville2009}.
	
	Thus, in the following we consider only the two stellar population labels: metallicity [Z/H] and [$\alpha$/Fe] enhancement.
	The statistical uncertainties on these quantities were estimated by \citet{saglia2018} from the relevant $\chi^2$ distributions, separately for the upper and lower limits.
	Initially, as a statistical uncertainty we took the larger of the two.
	We also calculated a local uncertainty, i.e.\ for each Voronoi cell we computed the standard deviation of the distribution of its neighbours.
	Finally, we derived an asymmetry uncertainty. For each pixel of a cell located at $(R_x,R_y)$ we found the value of the parameter at $(-R_x,-R_y)$ if it existed.
	We averaged those two values and took half of the difference between the given cell and its reflection as the asymmetry estimate.
	In the end, as the final uncertainty we took the largest of the three estimates. 
	
	We converted the available data to a square grid on the plane of the sky, which we will use in our modelling.
	\citet{saglia2018} made available the original positions of the VIRUS-W fibres and their allocation to the Voronoi cells.
	Since we needed to divide the sky plane into cells corresponding to the spectra, we implemented the following procedure.
	First, we divided the plane of the sky into a fine grid of square pixels of $1\arcsec$ size.
	Then, for each pixel we took the data values of the closest observed fibre, provided that it was closer than $5.35\arcsec$.
	This value ensures that all pixels inside the triangular fibre pattern of VIRUS-W are uniquely assigned.
	If for a given pixel the closest fibre is farther away, we treated that pixel as missing and we do not further use it in our considerations.
	
	\section{Methods}
	\label{sec_methods}
	
	In this section we describe the modelling technique we use in this work.
	First, we introduce our made-to-measure procedure and the assumptions it relies on.
	Next we test it on a set of mock data, and then
	discuss related conceptual issues. 
	Finally, we describe how our technique is applied to M31 using the dynamical model of \citet{blana2018}.
	
	\subsection{M2M modelling}
	
	Our aim in this contribution is to construct an $N$-body model of a stellar populations parameter $\phi$, for example metallicity, $\alpha$-enhancement or age.
	To achieve this, we use an observed map of the mean of the given parameter.
	Such a measurement is believed to be robustly obtained from full spectral fitting \citep[e.g.][]{cid_fernandes2013, cid_fernandes2014} or Lick indices \citep{trager_somerville2009}.
	For each line of sight $j$ (e.g.\ a pixel or a Voronoi cell) we use a mass-weighted mean $\Phi_j^{\mathrm{D}}$ and its associated uncertainty $\sigma_{\Phi,j}$.
	
	Furthermore, let us assume that we already have a dynamical, equilibrium $N$-body model of the galaxy, obtained through fitting the surface density and the kinematic data.
	We are keeping the dynamical model fixed and we are \emph{not} using the stellar populations to alter the model.
	In other words, in the M2M context, we are basically keeping the particle mass weights constant.
	This is an important point, since we will require our stellar population model to be consistent with the orbital distribution, therefore we will partially lift some of the degeneracy related to a deprojection of an image into a fully three-dimensional distribution.
	
	Now let us assign to every particle $i$ a single value of the parameter of interest $\phi_i$ (e.g. [Z/H], [$\alpha$/Fe] or $\log$ age).
	We then \emph{observe} our model galaxy from the same distance, at the same viewing angles and through the same lines of sight as the  real galaxy we are considering.
	For each line of sight $j$ we calculate the model observables as
	\begin{equation}
		\Phi_j^{\prime\mathrm{M}} = \frac{\sum\limits_{i\in j} m_i \phi_i}{\sum\limits_{i\in j} m_i},
	\end{equation}
	where $m_i$ are the masses of the particles and the sums are performed over all the particles present along the line of sight $j$. 
	Such a measurement might be quite noisy, thus we replace it by a time-averaged value, as originally proposed by \citet{syer_tremaine1996}
	\begin{equation}
		\Phi_j^{\mathrm{M}}(t) = \frac{1}{\tau}\int\limits_{0}^{\infty} \Phi_j^{\prime\mathrm{M}} (t-t') \exp\left(-\frac{t'}{\tau}\right) \mathrm{d}t',
		\label{eq_smth}
	\end{equation}
	where $\tau$ is a constant chosen in relation to the dynamical timescale.
	In practice, we approximate the integral by a discrete rule, which updates the value after each iteration \citep[see][]{syer_tremaine1996, delorenzi2007}.
	
	The value ''observed'' in the model is then compared with the data using a $\chi^2$ statistic
	\begin{equation}
		\chi^2 = \sum\limits_{j} \frac{\left(\Phi_j^{\mathrm{M}} - \Phi_j^{\mathrm{D}}\right)^2}{\sigma_{\Phi,j}^2},
	\end{equation}
	where the summation goes over all of the observed lines of sight.
	
	In the usual made-to-measure manner, we want to change the particle values of the parameter $\phi_i$ so that they fit the data optimally.
	We construct a merit function $F=-\frac{1}{2}\chi^2$ and while the particles orbit in the galactic potential, we apply the following force-of-change
	\begin{equation}
		\frac{\mathrm{d}\phi_i}{\mathrm{d}t} = \epsilon \frac{\partial F}{\partial \phi_i},
		\label{eq_foc}
	\end{equation}
	where $\epsilon$ is a suitably-chosen numerical parameter.
	In practice, we apply this equation, using the Euler method, in regularly spaced intervals, which we call \emph{iterations}.
	Hence, Eq. \eqref{eq_foc} can be understood as using the gradient ascent method to maximise $F$.
	In the dynamical formulation of M2M it is common to use an entropy term in the merit function \citep{syer_tremaine1996} in order to reduce the width of the particle mass distribution.
	However, it is natural to expect a non-negligible width and skewness of the distribution of stellar properties.
	Hence, we decided not to include any entropy term.
	
	The possible values of $\phi_i$ that a particle can have should be limited for both physical reasons and due to limitations of the technique used to obtain the data.
	For example, one would limit ages to values of $0$--$13.8$ Gyr, while metallicity and abundance would be limited by the extent of the stellar library.
	
	As in \citet{blana2018}, we calculate the potential using the hybrid method of \citet{sellwood2003}.
	We combine a polar-grid solver of \citet{sellwood_valluri1997} (updated by \citealt{portail2017a} to accommodate different softening lengths in the radial and vertical directions) and a spherical harmonic solver of \citet{delorenzi2007}.
	We let the potential rotate around the minor axis of our model with the angular velocity equal to the pattern speed of the bar.
	In such a potential we evolve the positions and the velocities of the particles, effectively treating them as test particles.
	
	The fitting procedure follows a usual M2M route.
	First, we initialise all of particle $\phi_i$.
	Then, we let the model evolve for $N_\mathrm{smooth}$ iterations, so that the observables are properly smoothed.
	Next, for $N_\mathrm{fit}$ iterations we fit $\phi_i$ of the particles, following \eqref{eq_foc}.
	Finally, we let the model relax for $N_\mathrm{relax}$ iterations, so we can check if it was not overfitted.
	From the final values of the particle $\phi_i$ we compute other interesting characteristics, such as profiles and deprojected maps.
	
	Our method can be summarised as follows.
	We start with an $N$-body model that is a faithful representation of a galaxy.
	We tag every particle with a single value of e.g.\ metallicity.
	As the particles move in the galactic potential, we adjust their metallicities to fit the observed map of the mean metallicity in the galaxy.
	
	We made a number of tests of the method, also with moderately inclined mock galaxies (i.e.\ not edge-on).
	We found that if we used the simplest initialisation of $\phi_i$, making it equal to a constant value everywhere, our technique was not able to recover on its own the correct vertical gradient of $\phi$. 
	This is related to deprojection degeneracies; see e.g., Fig.~16 of \citet{zhu2020}, and is here discussed further in Section~\ref{sec_mock_test} and Appendix~\ref{app_vertical_prior}.
	To improve on this issue, we initialise the $\phi_i$ values of the particles depending on height above the galaxy plane, according to
	\begin{equation}
		\label{eq_init}
		\phi_i(z_i) = G (|z_i| - z_0) + N,
	\end{equation}
	where $G$ and $N$ are constants, $z_i$ is the particle's vertical coordinate and $z_0$ is a normalisation constant, equal in our case to the mean absolute vertical coordinate of all particles (which is equal to the scale-height in the case of the exponential profile).
	We try a set of possible $G$ and $N$ and check which one results in the lowest value of final $\chi^2$.
	Then we use this initial condition for final results and uncertainty estimation.
	One could wonder if a linear function of the vertical coordinate is sufficient, or should we use a different, possibly more complicated function.
	Unfortunately, we are not aware of any observationally or theoretically motivated functional form for the vertical profiles of metallicity or $\alpha$-enhancement.
	We decided to use the next simplest polynomial (after a constant value) of degree one, which has two free parameters.
	As we validate in Sect. \ref{sec_mock_test}, in this way we are able to capture most of the variation, but not small details.
	Another scheme may be better, but one would need more data to judge it.
	
	To estimate the uncertainties of e.g.\ profiles, we use the following procedure.
	We initialise the particle $\phi_i$ using the best-fit vertical profile with additional Gaussian noise added to seed randomness.
	To the data values $\Phi_j^\mathrm{D}$ we add Gaussian noise with zero mean and standard deviation of $\sigma_{\Phi,j}$.
	Next, we refit these new data and recompute the quantities of interest, e.g. the profiles.
	We repeat this procedure $100$ times and from the variance of the profiles we estimate their uncertainty.
	To such a statistical uncertainty we add in quadrature a spread of the profiles that were obtained from models with different initial vertical profiles and were within 1$\sigma$ from the $\chi^2$ minimum.

	\subsection{Tests on mock data}
	\label{sec_mock_test}
	
	When a new method is proposed it should be verified on suitable mock data so one can be reasonably convinced that it gives correct answers.
	Hence, here we describe our tests.
	
	To create mocks we used the chemodynamical barred galaxy model created by \citet{portail2017b}.
	It consists of $10^6$ stellar particles in dynamical equilibrium with its dark matter halo.
	Originally, it was a disc galaxy with a bar of $5$ kpc length.
	Since we wanted to make a comparison to M31, which has a $4$ kpc bar, we decided to adjust the extent of the model.
	We scaled all of the sizes by a factor of $4/5$, all of the velocities by also $4/5$ and masses by $(4/5)^3$
	\footnote{
		Recall that in dynamics there are three basic dimensions, which can be chosen as e.g. length, velocity, and mass.
		The gravitational interactions are invariant under the transformation $x\to \alpha x $, $v\to \beta v$ and $m\to \alpha\beta^2 m$, where $x$ denotes coordinates, $v$ denotes velocities and $m$ denotes masses.
		Note that due to this transformation time $t\to (\alpha/\beta)t$.
	}.
	We observed the model at the distance ($785$~kpc), the inclination ($77\degr$), and the position angle of the bar with respect to the line of nodes ($54.7\degr$), the same as in M31.
	
	In \citet{portail2017b} each particle has four weights, representing fractions of the particle mass corresponding to four bins of the stellar [Fe/H].
	For the purpose of this test we assign each particle a single mean metallicity that reflects the fractional weights.
	Thereby, we obtain a reasonable model of mean metallicity in a barred galaxy.
	We use the same set of the Voronoi line-of-sight cells as \citet{saglia2018} and we let the model evolve for $10^4$~it ($1\ \mathrm{it}=1.1\times 10^{-4}$ Gyr) to smooth the observables.
	In order to create a realistic observed [Fe/H] map we added Gaussian noise with zero mean and $\sigma_\mathrm{[Fe/H]}=0.04$ dex, equal to the average uncertainty of [Z/H] reported by \citet{saglia2018}.
	
	We limit the possible values of [Fe/H] to the same range as the [Z/H] grid of models in \citet{saglia2018}, i.e.\ from $-2.25$ to $0.67$ dex.
	As the underlying dynamical model we use the rescaled model from \citet{portail2017b}, thus we do not have any additional uncertainty that would arise if the model used in the fitting did not correspond to the density distribution of the "data".
	How important this uncertainty is depends strongly on how tightly the dynamical model is constrained in the case at hand. Therefore we evaluate its impact on the recovered metallicity profiles in M31 directly in Section \ref{sec_met}, using the set of models available for Andromeda.
	
	\begin{figure}
		\centering
		\includegraphics{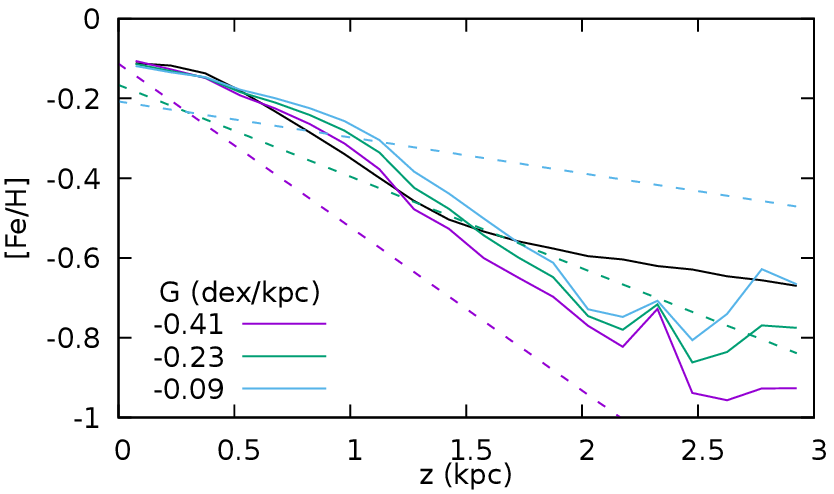}\\
		\includegraphics{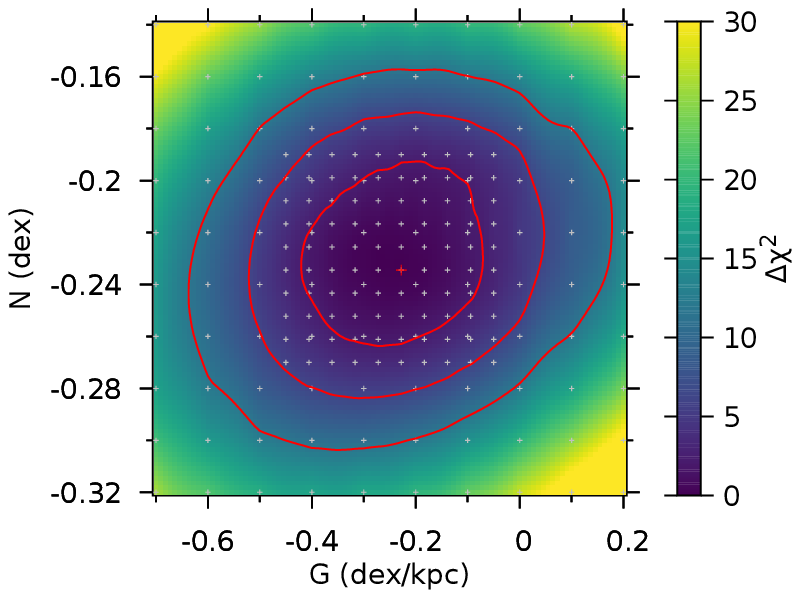}
		\caption{
			Top panel:
			initial (dashed lines) and final (solid lines) vertical metallicity profiles of the mock galaxy.
			The green line marks the model with the lowest final $\chi^2$, while the blue and violet lines correspond to 1$\sigma$-worse models.
			The black line shows the vertical profile of the original mock galaxy.
			Bottom panel:
			map of final $\Delta \chi^2$ as a function of initial $G$ and $N$ in Eq.~\eqref{eq_init}.
			Grey crosses indicate the actually computed models, while the underlying coloured map is a result of interpolation.
			Red contours depict 1-, 2- and 3-$\sigma$ regions.
			The red cross indicates the model with the lowest $\chi^2$.
		}
		\label{fig_mock_vertical}
	\end{figure}

	We illustrate our procedure of fitting the vertical profile in the top panel of Figure \ref{fig_mock_vertical}.
	We start with the initial vertical metallicity profiles (priors), shown by the dashed lines.
	It may seem that they are far away from the original vertical metallicity profile of the mock galaxy (black solid line).
	However, in the initial smoothing phase the gradients get shallower, due to phase mixing of the particles.
	Moreover, the prior is most important where the data constraints are weak, i.e. at large heights and large distances from the centre.
	After running the modelling code, each initial profile results in a~slightly different final one (solid lines) and a different final value of $\chi^2$.
	In Figure \ref{fig_mock_vertical} we show models with different initial gradient $G$, but the same normalisation $N$ at $z_0=\langle|z|\rangle=0.297$~kpc.
	The initial linear profiles are transformed into more complicated functions.
	
	The best fit to the mock galaxy profile (i.e. the lowest $\chi^2$) is obtained for an initial $G=-0.23$ dex/kpc , while the other two models correspond to 1$\sigma$ worse cases; see lower panel of Figure \ref{fig_mock_vertical}.
	The blue line in fact approximates a constant initial prior.
	The largest difference between the best-fit profile and the other models within the 1$\sigma$ region is considered as a part of the final uncertainty.
	In the bottom panel of Figure \ref{fig_mock_vertical} we show a $\Delta\chi^2$ map of the models as a function of $G$ and $N$.
	We stress again that the exact values of $G$ and $N$ are not directly related to the final vertical profile.
	
	We checked the impact of the vertical prior on the radial profile and it turned out that in the central part it is rather minor.
	In the outer part it is more noticeable, which is reflected in a wider uncertainty band for the radial profile beyond $R\sim7$ kpc. 
	In all these tests, the prior did not depend on radius. In Appendix~\ref{app_radial_prior} we show the effect of initial priors with a radial gradient. We find that the optimal radial gradient is consistent with zero, and varying it within the $1\sigma$ region has minor impact on the recovered radial profile.
	
	\begin{figure}
		\centering
		\includegraphics{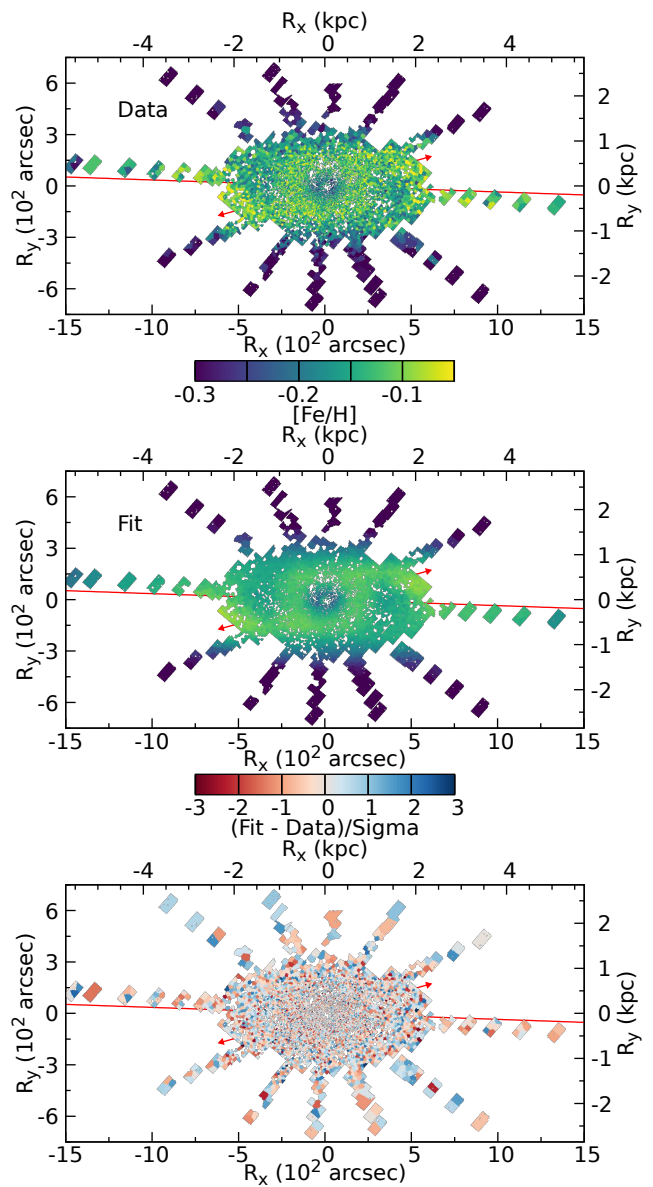}
		\caption{Top: mock map of mean [Fe/H] with a noise of $\sigma_\mathrm{[Fe/H]}=0.04$~dex.
			Middle: best M2M dynamical model.
			Bottom: standardised residuals.
			The solid red line marks the position angle of the projected disc major axis, while the red arrows mark the orientation and the extent of the bar.}
		\label{fig_mock_maps}
	\end{figure}
	
	In Figure \ref{fig_mock_maps} we present the mock data, the map of mean metallicity from the best model, and the standardised residuals of the fit.
	The reduced $\chi^2_\mathrm{red} = 0.99$ indicates an almost perfect fit, which can be also glanced from the residual plot.
	However, such a good result is not surprising since the dynamical model is correct and the uncertainties of the data are perfectly Gaussian, with known amplitudes.
	We conclude that our procedure does not have any persistent problems fitting major features of the mock data.
	
	\begin{figure}
		\centering
		\includegraphics{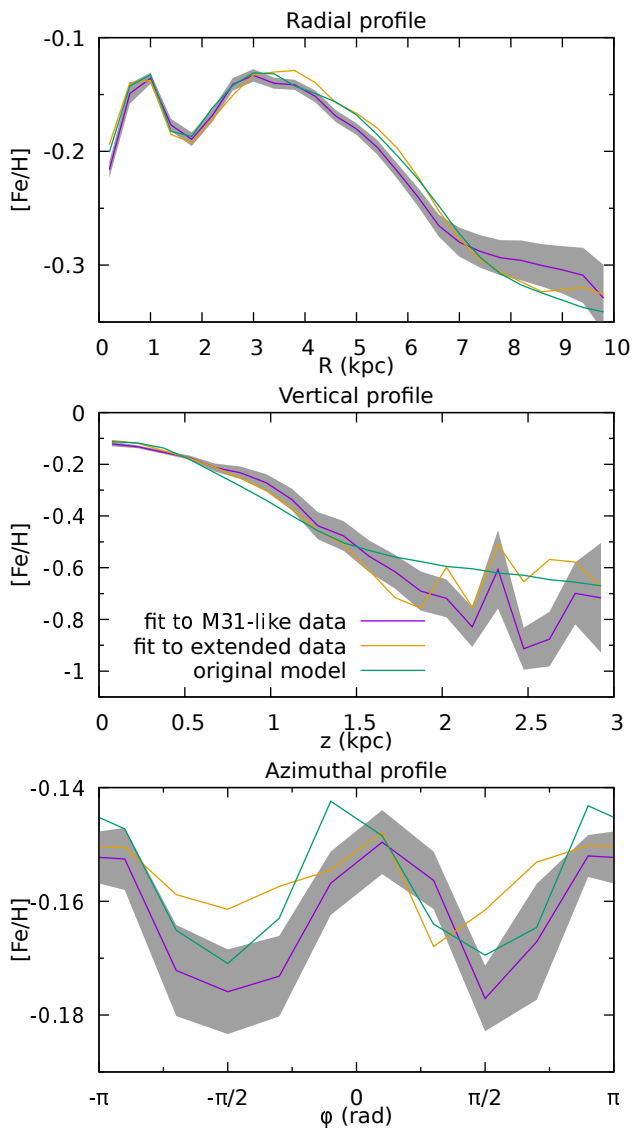}
		\caption{Metallicity profiles calculated from the model fitted to M31-like mock data (violet, with a grey band of uncertainty), compared to profiles of the original model (green).
			The yellow lines show profiles resulting from a model fit to spatially-extended mock data.
			From top to bottom: radial profile (as a function of cylindrical radius), vertical profile and azimuthal profile.
		}
		\label{fig_mock_profiles}
	\end{figure}
	
	Let us now compare results of the fitting with the original model.
	In Figure \ref{fig_mock_profiles} we compare various profiles computed both from the original model of \citet{portail2017b} and our best-fitting M2M model.
	In the top panel we plot the [Fe/H] profile as a~function of cylindrical radius $R$.
	We took into account all stellar particles with $|z|<3$ kpc.
	In the second panel we plot the vertical profile for all particles $R<5$ kpc.
	In the last panel we present the azimuthal profiles for the particles with $R<5$ kpc and $|z|<3$ kpc. 
	
	Our modelling routine is able to recover all of the features of the radial profile, including the [Fe/H] peaks at $\approx 1$ kpc and close to the end of the bar at $\approx 3$ kpc, as well as a further negative gradient up to 10 kpc.
	At first sight this might be surprising, since the data along the major axis reach only 5 kpc, however the key lies in the fields along the minor axis, which reach large distances due to the projection at the inclination of $77\degr$.
	In the azimuthal profile, the main variation is related to the enhancement along the bar, which is aligned with $\varphi=0,\,\pm\pi$.
	
	As we could already see in Figure \ref{fig_mock_vertical}, the vertical profile is not reproduced exactly.
	However, we are able to recover an average vertical gradient.
	One can also infer a certain degree of non-linearity in the profile, which is flatter closer to the galaxy plane than at larger heights. 
	In Appendix~\ref{app_vertical_prior} we show similar results as in Fig.~\ref{fig_mock_profiles} but for a lower inclination model ($i=45^\circ$). Deprojection degeneracies are stronger in this more face-on case, as is apparent particularly in the larger deviations of the recovered vertical profiles from the input model profile.
	
	We made another test to judge if the vertical profile is influenced by limited on-sky data coverage.
	We constructed a complete data set similar to Fig.\ \ref{fig_mock_maps}, filling the whole 6 by 12 kpc field.
	To construct the Voronoi tessellation we used the package \textsc{VorBin} by \citet{cappellari_copin2003}.
	However, for technical reasons\footnote{\textsc{VorBin} did not converge for a finer grid.} we used four times fewer Voronoi cells.
	The yellow lines in Fig.\ \ref{fig_mock_profiles} depict the result of modelling these data.
	The vertical profile is closer to the original model (green line) especially for $z>2$ kpc.
	The radial profile provides a slightly worse match at $R\approx 4$ kpc, but much better for $R>4.5$ kpc, and the azimuthal profile remains with similar deviations from the true profile.
	
	Some discrepancy therefore remains even for spatially complete data. This could be ascribed to other reasons, such as the orbital degeneracy discussed in the next subsection, the lower spatial resolution, or the noise in the data.
	It may be surprising that the recovery of the vertical profile in case of an \emph{almost} ($77\degr$) edge-on galaxy is problematic.
	However, recall that our earlier tests showed that the constraints on the vertical profiles in more face-on galaxies are much weaker.
	Furthermore, the vertical variation leaves rather small imprints in an on-sky map, which can be dwarfed by radial changes.
	
	Our estimate of the uncertainties seems to be justified in that the recovered profiles generally are no farther than $1$--$2\sigma$ away from the true profiles.
	On the other hand, the uncertainty in the vertical profile at small heights ($z\lesssim 1$ kpc) appears to be underestimated.
	We note that errors within a single profile and between different spatial profiles are highly correlated.
	First, the model is subject to degeneracy due to the galaxy being projected on the sky. Secondly, as building blocks of our M2M model we are using spatially-extended orbits, which contribute in many different locations.

	\subsection{Discussion of the method}
	\label{sec_discuss_method}
	
	One could ask the question what we can actually constrain with this method, using maps of mean $\Phi^\mathrm{D}$ value on the sky.
	First, let us recall how one may understand M2M dynamical models, which we use as a basis for our work.
	They are composed of a~set of particles in a dynamical equilibrium.
	However, they can be also regarded as a collection of \emph{orbits}.
	Each orbit has an associated mass (equal to the mass of the particle) and the sum of the orbital masses gives a density distribution, whose potential generates the aforementioned orbits.
	This interpretation resembles the \citet{schwarzschild1979} modelling, which was used by \citet{long_mao2018}, \citet{poci2019} and \citet{zhu2020} to study stellar populations.
	We note that \citet{long2016} implemented a similar technique, however it was applied to absorption line strengths, limited to symmetrised data and not tested besides converging to an acceptable $\chi^2$.
	
	If we regard our system as a collection of orbits $\{i\}$, then we are actually fitting the mean values of a population parameter on a given orbit.
	However, we \emph{cannot} constrain a distribution function of the parameter on the orbit, e.g., whether it is $\delta$-like or very broad.
	Therefore, we also cannot derive the distribution function of $\phi$ for the whole galaxy, we can only give a rough lower limit on its width.
	As an extreme example, one can imagine that a spatial map of a galaxy is constant everywhere and equal to $\phi_0$.
	One would conclude that the mean value on all of the orbits is also equal to $\phi_0$.
	However, in reality, on each orbit there might be two stars, one having $\phi=\phi_0+\Delta$  and the other one with $\phi_0-\Delta$.
	Using only the map of the mean value one cannot determine $\Delta$.
	Having constrained the mean values of $\phi$ on the orbits, we can then project these quantities as a function of more accessible variables, such as spatial coordinates or velocities.
	In particular, in this contribution we are focusing on profiles as a~function of position, and deprojected maps, but we could also plot $\phi$ as a~function of velocity coordinates or actions.
	
	Thus, our method, or any similar method using the same type of data, naturally produces a distribution function of $\phi$ at a given position, which reflects the distribution of the $\phi$ values on the orbits that pass through this position, but it does not necessarily reflect the true distribution of $\phi$ there.
	Some progress could be achieved by adding physically-motivated assumptions; for example, \citet{zhu2020} employed a prior based on an age-orbital circularity relation.
	In the present context, a natural choice would be a prior relating [Z/H] and [$\alpha$/Fe]; however, our M31 data show little relation between both quantities. 
	The other, preferred way to improve the recovery of the true distribution of stellar labels would be to use more data constraints such as from full spectral fitting \citep[see e.g.][Fig. 1]{peterken2020}.
	
	A further natural question could be asked about possible degeneracies between stellar orbits.
	Our method is in fact based on matching the projected surface density of $\phi$-values on the particle orbits to the data values.
	If a projection of an orbit could be linearly decomposed into projections of other orbits, then our technique would not be able to unambiguously assign stellar population labels to them.
	In principle, one expects such an occurrence, since the orbital phase space can be labelled by three actions \citep[see e.g.][]{binney_tremaine2008}, while our data is inherently two-dimensional.
	
	It is relatively easy to understand the issue for axisymmetric disc galaxies.
	In this case the orbits can be classified based on three conserved actions.
	The angular momentum $L$ (equivalent to the azimuthal action) describes the size of the orbit.
	The radial action $J_r$ refers to the orbital eccentricity.
	The vertical action $J_z$ describes the vertical thickness of the orbit.
	First, let us consider only planar orbits (i.e.\ $J_z=0$).
	Then the projected density distribution of an orbit with non-zero $J_r$ can be constructed as a sum of density distributions of circular orbits.
	However, if we project a vertically extended, axisymmetric orbit (with non-zero $J_z$) at an intermediate inclination, it will appear more extended along the minor axis of the galaxy than a planar orbit of the same radial extent.
	Therefore it appears that the $J_z$ dimension is independent of the other two.
	
	The case of barred (i.e.\ non-axisymmetric) galaxies is more complicated.
	On the one hand one still expects some degeneracy because of the different number of dimensions of the phase-space and the data, respectively.
	On the other hand, neither the classical planar orbital families x$_1$--x$_4$ \citep{contopoulos_papayannopoulos1980}, nor the vertically extended orbital families \citep{skokos2002}, nor the orbits from the actual $N$-body simulations \citep[e.g.][]{valluri2016, gajda2016} exhibit obviously degenerate orbital projections.
	This issue certainly warrants further investigation.
	
	Also some of these degeneracies would be reduced with more detailed data, or lacking this, if additional priors were included in the modelling.
	As mentioned above, such priors could include relations between the stellar population labels themselves, e.g. metallicity and alpha, or when available, age, or priors with an explicit dependence on the orbital parameters, such as circularity or actions.
	The drawback with this approach is the difficulty of distinguishing what is the prediction of the model and what is just a corollary of the assumed relations.
	
	The practical conclusion from our mock tests, and similarly from \citet{zhu2020}, is that despite the remaining degeneracies there is considerable information one can derive from spatially resolved mean maps of stellar population parameters.
	We showed that the radial and azimuthal profiles are well-constrained, and in nearly edge-on systems one can retrieve also the vertical profile when applying some extra care.
	
	\subsection{M31 dynamical model and parameters used in stellar population modelling}
	
	As the basis for our stellar population fitting we employed the JR804 $N$-body model of the Andromeda Galaxy constructed by \citet[see Tab. 1]{blana2018}.
	It consists of $2\times 10^6$ dark matter particles with Einasto density profile, $10^6$ disc particles (which include the bar and its box/peanut bulge) and $10^6$ classical bulge particles.
	Using the M2M technique it was fitted to the moments of the velocity distribution derived by \citet{opitsch2018} and
	to the $3.6$ $\muup$m-band surface brightness maps from the Spitzer Space Telescope obtained by \citet{barmby2006}.
	
	\citet{blana2017,blana2018} concluded that M31 is a~barred galaxy.
	They found that the bar position angle (in the plane of its disc) with respect to the line of nodes is equal to $54.7\degr$.
	The bar rotates with a pattern speed of $\Omega_p = 40\pm5$ $\mathrm{km}\,\mathrm{s}^{-1}\,\mathrm{kpc}^{-1}$ and has length of $\approx4$ kpc.
	The dark matter halo was found to follow an Einasto profile with a mass of $1.2^{+0.2}_{-0.4}\times 10^{10}$ M$_{\odot}$ within $3.2$ kpc and a stellar mass-to-light ratio in the $3.6$ $\muup$m band of $\Upsilon_{3.6\muup\mathrm{m}}=0.72\pm0.02$ $\mathrm{M}_\odot\,\mathrm{L}_\odot^{-1}$, assumed to be a single constant.
	\citet{blana2018} in their paper provide a set of models that fit the data well and which could be thought of as "1$\sigma$ models".
	We use those models as a basis for the computation of the uncertainty induced by the variation of the dynamical model.
	
	The \citet{blana2018} M2M model of the Andromeda Galaxy was fit to a range of observational data and reproduced them correctly.
	While it did not explicitly fit the vertical scale height, it was based on a model survey by \citet{blana2017}, who tested various configurations.
	Hence, the model's mean scale height of $0.72$ kpc is comparable to the $h_z=0.86\pm0.01$ kpc that can be inferred from the PHAT survey (\citealt{dalcanton2015}, see also \citealt{bhattacharya2019}).
	\citet{blana2018} also included a dust model to screen material behind the disc plane of M31.
	
	Having a detailed dynamical model is an excellent basis for the modelling in this paper since the stellar orbits are thereby determined consistently with the photometry and kinematics. However, some uncertainties remain.
	We assume that the population labels are mass-weighted and, hence, their best-fit distribution depends on the density distribution.
	\citet{blana2018} fitted the $3.6$ $\muup$m IRAC image that traces the old giant stars (the bulk of the population).	While the fitting was luminosity-weighted, the assumption of using a single constant mass-to-light ratio implies a trivial relation between particle masses and their $3.6$ $\muup$m luminosity weights, which, in turn, implies that both mass- and light weighted averages give exactly the same result.
	
	However, the measured kinematics are light-weighted in the V band where younger or more metal-poor stars could have a~bigger impact in some regions \citep[see e.g.][]{portaluri2017}.
	Thus, while the surface mass distribution is well-constrained by the infrared photometry, the distribution of the particle orbits is biased towards the V band kinematics without considering the V band photometry. This could be a significant effect in the disk regions with more recent star formation.
	A future M2M model might be improved by fitting the V-band simultaneously with the IRAC $3.6$ $\muup$m to better model the kinematic structure in these regions.
	
	An observant reader would notice a small inconsistency in our approach. The dynamical model of \citet{blana2018} assumes a constant $M/L$, whereas here we determine a posteriori a distribution of metallicities and $\alpha$-enhancements which should lead to slight variations of the $M/L$ between different parts of the galaxy. 
	The reason not to vary $M/L$ was our uncertainty whether the stellar population labels are precise enough to constrain the dynamics.
	Additionally, the projected variability of [Z/H] is of the order of 0.2 dex, which implies a difference in $3.6\muup$m-band $M/L$ of about $5\%$ \citep{meidt2014} and about $10\%$ in V-band \citep{vazdekis2010}, both of which are of the same order as the ${\sim}\,4\%$ uncertainty of our model's $M/L_{3.6\muup\mathrm{m}}$ (including a systematic uncertainty from different choices of the dark matter profile).
	We conclude that the impact of a variable $M/L$ is rather small and does not influence our main results.
	
	As a side note, we remark that to date none of the dynamical models of stellar populations in galaxies is proven to be fully internally self-consistent.
	Concerning the model of \citet{poci2019}, the remaining question is whether the mass distribution inferred from the orbital light-weights and their respective mass-to-light ratios reproduces the deprojected stellar mass distribution used to generate the orbits.
	\citet{portail2017a}, \citet{poci2019} and \citet{zhu2020} use a single number to convert from an observed quantity to mass ($M/L$ in \citealt{poci2019} and \citealt{zhu2020}; "mass-to-clump ratio" in \citealt{portail2017a}).
	It appears that making an actual \emph{fully} self-consistent model is a worthwhile goal for a future research. 
	This should not be too difficult in M2M: during the iterative joint modelling of the dynamics and the stellar population parameters, the $M/L$ of all particles would be adapted on the fly according to their current ages, metallicities and $\alpha$ enhancements.
	However, this would require significantly more detailed data than available here.
	
	Following \citet{blana2018}, we set 1 iteration to $1.18\times 10^{-4}$ Gyr.
	Similarly, we fix the smoothing time scale to $\tau=1.6\times10^3$ it, which corresponds to the orbital timescale at $R=5$~kpc.
	We tested different values of $\epsilon$ that controls the strength of the force-of-change in \eqref{eq_foc}.
	We found that in the case of [Z/H] the best results (i.e. the lowest $\chi^2$) are obtained for $\log_{10}\epsilon=-1.9$ and in the case of [$\alpha$/Fe] for $\log_{10}\epsilon=-2.1$.
	We smooth the observables initially for $3\times 10^3$ it, we fit for $50\times10^3$~it, until $\chi^2$ converges to a constant value and then we let the system relax for $7\times10^3$ it.
	As usual in the M2M modelling, after we finish fitting $\chi^2$ increases in the relaxation phase, by about $\Delta \chi^2_\mathrm{red}\sim0.1$, and stabilises at a new and final value.
	
	\section{Results}
	\subsection{M31 metallicity}
	\label{sec_met}
	
	\begin{figure}
		\centering
		\includegraphics{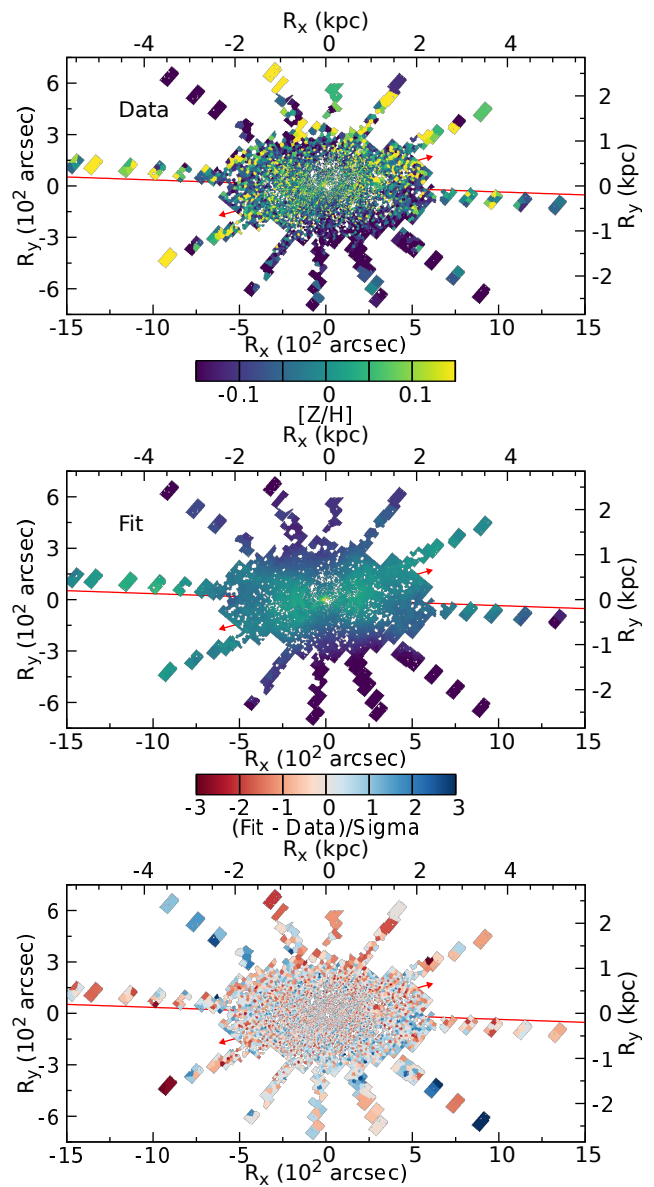}
		\caption{Metallicity maps of the Andromeda Galaxy.
			Top: map of the measured [Z/H].
			Middle: the best model.
			Bottom: standardised residuals.
			The solid red line marks the position angle of the projected disc major axis, while the red arrows mark the orientation and the extent of the bar.}
		\label{fig_met_obs}
	\end{figure}
	
	First we present our modelling of the [Z/H] distribution in Andromeda.
		Similarly as in the mock test, we use a vertical initial prior with parameters $(G,N)$, finding that $G=-0.21\pm0.27$ dex/kpc and $N=-0.02\pm0.16$ dex result in the models with lowest overall $\chi^2$.
	In Figure \ref{fig_met_obs} we show the data (top), the best model (middle) and the standardised residuals (bottom).
	The white spaces correspond to the lines-of-sight not covered by the data.
	The $(R_x,R_y)$ reference frame \citep[the same as in][]{blana2018} is rotated by $50\degr$ clockwise with respect to the sky coordinates, so that the projected major axis of the M31 disc is almost aligned with the $R_x$-axis.
	The kpc labels correspond to the sizes on the sky at the distance of M31.
	To convert $R_y$ into a distance in the plane of the disc, one should multiply it by $(\cos i)^{-1}\approx4.4$; thus the elliptical region on the sky covered densely by the data fields extends to 5.5 kpc along the minor axis.
	We also note that the data cover nearly the entire bar length.
	The possible values of [Z/H] are limited to the range -2.25 to 0.67 dex, as in \citet{saglia2018}.
	
	Overall, the data are fitted very well by the model ($\chi^2_\mathrm{red} = 0.94$), especially in the central parts.
	However the fit appears to be worse for some of the high values in the spokes ($R_y>300\arcsec$) covering the disc, where [Z/H] is underestimated.
	The data itself exhibit a significant asymmetry, with the top part of M31 having higher metallicity than the bottom part (see top panel of Fig. \ref{fig_met_obs}).
	This could be related to dust obscuration by the disc whose nearby parts are in front of the bulge at $R_y>0$ \citep[see][Sect. 3.2.3]{blana2018}.
	In Appendix \ref{app_dust} we discuss in more detail the arguments that the dust is at least partially responsible for the asymmetry.
	
	\begin{figure}
		\centering
		\includegraphics{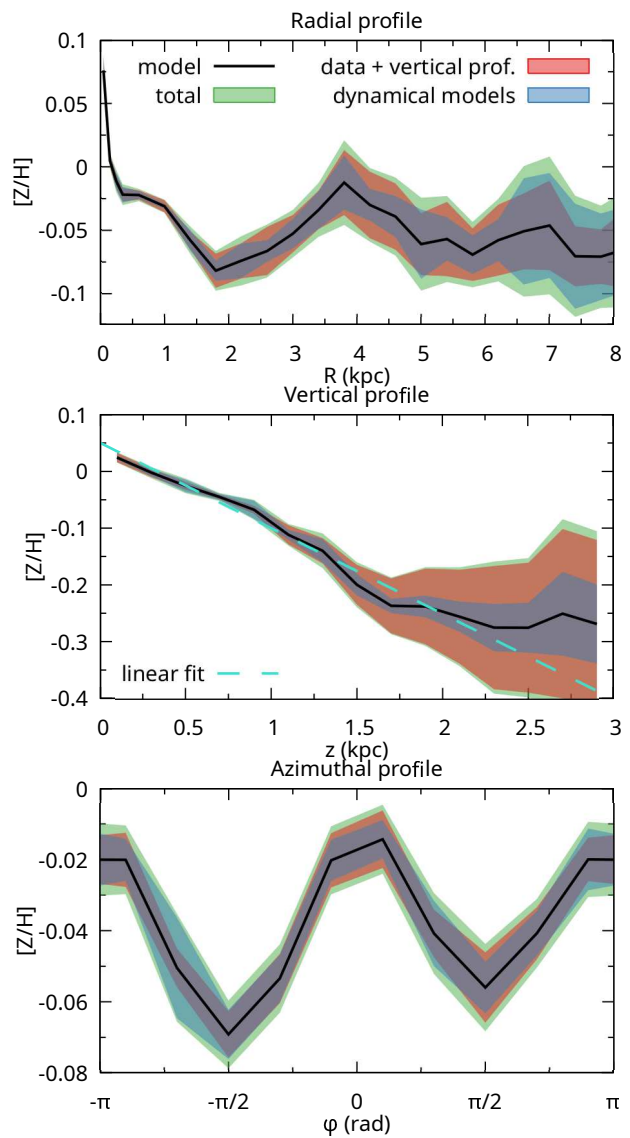}
		\caption{Metallicity profiles calculated from the best model.
			From top to bottom: the radial profile (as a function of the cylindrical radius), the vertical profile, and the azimuthal profile (black lines).
			The coloured bands show the different uncertainties.
			In red we mark the sum of the errors stemming from the uncertainties in the data and the vertical profiles, in blue we show the error range resulting from the uncertainty of the underlying dynamical model, and the green bands mark our final uncertainty estimates, which are the sum (in quadrature) of the three.
			Note that overlapping red, blue, and green appears as grey, and overlapping red and green as orange.
			In the middle panel the turquoise dashed line represents an uncertainty-weighted linear fit to the model curve.}
		\label{fig_met_profiles}
	\end{figure}
	
	In Figure \ref{fig_met_profiles} we present the metallicity profiles for M31, calculated from the best model.
	We plot [Z/H] as a function of cylindrical radius $R$ (i.e. measured in the plane of the disc), height above the disc plane $z$, and azimuthal angle $\varphi$.
	We took into account only the particles with $|z|<3$ kpc.
	For the vertical profile we used particles with $R<5$ kpc, and for the azimuthal profile we took into account only particles with $R<5$ kpc and $|z|<3$~kpc.
	
	For each profile we mark the different uncertainties. The red bands include the uncertainty stemming from re-fitting the model to different realisations of the data within their errors and from using different priors for the vertical profile within their respective 1$\sigma$ uncertainties. Blue bands show the uncertainty range from the dynamical model, derived from the 11 models in Table 1 of \citet{blana2018}, which the authors of this study deemed \emph{acceptable models}.
	Their dark matter halos all have the Einasto density profile, but they differ somewhat in the bar pattern speed, mass-to-light ratio and dark matter mass in the bulge region. 
	We refitted the M31 metallicity observations for all these models with the same initial vertical profile as in our fiducial case.
	We inspected the resulting models of the [Z/H] distribution and they were qualitatively similar to the case of the fiducial dynamical model.
	To assess the quantitative differences, we computed the dispersion of these profiles with respect to the fiducial model and plotted it as the blue bands in Fig. \ref{fig_met_profiles}.
	This uncertainty has a similar if slightly smaller magnitude as the other types of uncertainty, except at large heights where the uncertainty from obtaining the vertical profile dominates.
	Finally, the green bands in Fig. \ref{fig_met_profiles} depict the overall uncertainty of our model, where we added in quadrature all three sources of error.
	
	\begin{figure*}
		\centering
		\includegraphics{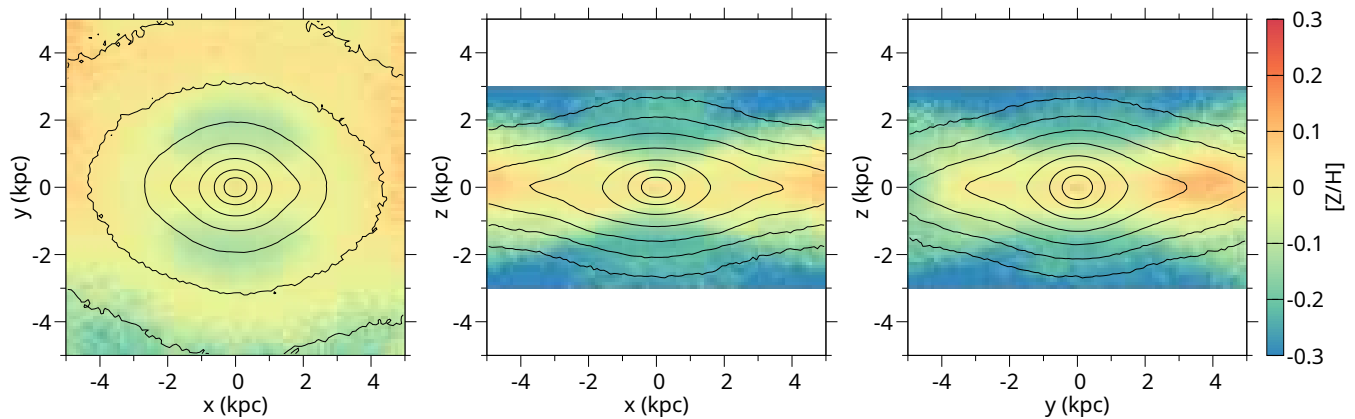}
		\caption{Deprojected maps of mass-weighted mean [Z/H] in M31. From left to right: face-on, edge-on and end-on view.
			Overplotted are surface density contours in a range $\log(\Sigma/(\mathrm{M}_\odot\,\mathrm{kpc}^{-2}))=8$--$10$ with a multiplicative step of $10^{1/3}$.}
		\label{fig_met_maps}
	\end{figure*}
	
	In the radial profile one immediately notices a spike in the central ${\sim}\,200$ pc, which is also discernible in the raw data of \citet[Fig.\ 22]{saglia2018} and was interpreted there as a sign of the classical bulge.
	Then, [Z/H] decreases further, up to $\approx 2$~kpc, which we interpret below in terms of the metallicity desert perpendicular to the bar.
	Further out, the azimuthally averaged profile starts to increase again, reaching a maximum around $4$~kpc from the centre.
	This coincides with the size of the bar in the model of \citet{blana2018}.
	Still further out, in the disc outside the bar, the profile decreases once again, but does not exhibit a steady negative gradient, possibly due to insufficient data coverage.
	We limit the radial range of the profile to $8$~kpc, because the asymmetry of the data (see Fig. \ref{fig_met_obs}) influences the results strongly beyond that point.
	Our profile can be compared to the radial metallicity profile obtained by the PHAT survey \citep{gregersen2015}.
	Their radial profile has a metallicity maximum at $R\approx 4.5$ kpc and further away [M/H] decreases linearly with radius.
	While we do not find a clear linear decrease beyond the end of the bar, our mean values are consistent with the median values reported in their Fig.~9.
	
	The azimuthal profile supports the conclusion by \citet{saglia2018} that [Z/H] is enhanced along the bar (aligned here with $\varphi=0,\,\pm\pi$).
	The [M/H] map by \citet{gregersen2015} also indicates that metallicity is enhanced along the bar, particularly close to its tip.
	
	The average vertical profile is consistent with linearity.
	Thus, we fitted a linear function using uncertainty-weighted least squares and obtained a vertical gradient of $\nabla_z\mathrm{[Z/H]} = -0.133\pm0.006$ dex/kpc (see Fig. \ref{fig_met_profiles}).
	Note that the uncertainty of the slope is likely underestimated, since it is driven by the well-constrained region at $|z|<1$ kpc.
	
	In Figure \ref{fig_met_maps} we show \emph{deprojected} maps of the mass-weighted mean metallicity.
	The minor axis of the disc is aligned with the \mbox{$z$-axis}, while the bar major axis is here aligned with the $x$-axis.
	To create these maps we used only the particles within $R<7.2$~kpc and $|z|<3$~kpc.
	Then, they were time-smoothed in the frame rotating with the bar using equation \eqref{eq_smth}, with a time-scale of $\tau=0.19$ Gyr.
	To guide the eye, the surface density contours are overplotted in black.
	
	In the central $3$ kpc of the face-on view one can see an enhancement along the bar and depressions in the direction perpendicular to it.
	Further away one can notice maxima at both ends of the bar, around $x\approx 4{-}5$ kpc.
	There is also a noticeable asymmetry between positive and negative $y$, from $|y|\sim4$ kpc outwards, with a larger metallicity inferred at positive $y$.
	This is a direct consequence of the top-bottom asymmetry of the data (Fig. \ref{fig_met_obs}).
	We checked that this feature is dynamically stable, continuing the relaxation phase for $5\times10^4$~it ($\approx 6$ Gyr).
	In Appendix~\ref{app_dust} we describe additional tests related to this asymmetry which indicate that orbits around the Lagrange points L$_4$/L$_5$ make it dynamically stable.
	Thus, in  principle, the [Z/H] distribution itself could by asymmetric in M31, e.g. due to enhanced star formation in spiral arms on one side of the galaxy.
	However, we cannot currently discriminate whether the inferred larger [Z/H] at large positive $y$ (see top panel of Fig. \ref{fig_met_obs}), are caused by the dust or whether they are due to intrinsic asymmetry of M31.
	Therefore we consider the bottom part of the map ($y<0$) more trustworthy.
	It exhibits a noticeable ring of enhanced metallicity, beyond which [Z/H] decreases outwards, compatible with \citet{gregersen2015}.
	
	\begin{figure}
		\centering
		\includegraphics{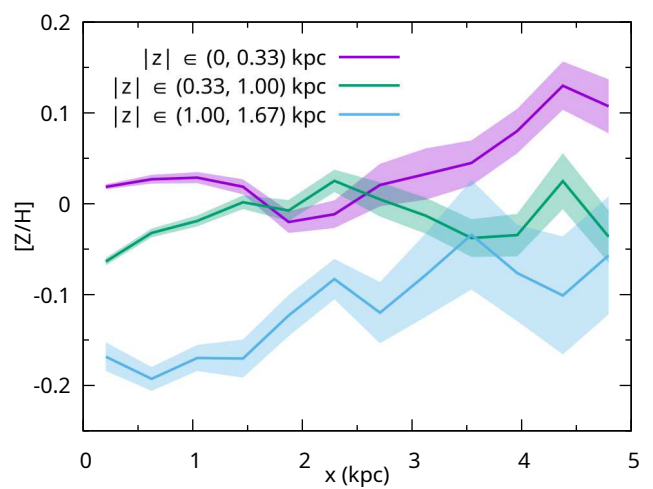}
		\caption{
				Metallicity profiles in three adjacent horizontal strips in the central region of M31.
				The b/p bulge extends to $x\approx 2.5$--$3$ kpc and $z\approx 1$ kpc, while the bar length is $\sim4$ kpc. The maximum along the second (green) [Fe/H] profile corresponds approximately to the end of the b/p bulge.
		}
		\label{fig_met_bp_profiles}
	\end{figure}
	
	The edge-on view clearly exhibits an X-shape.
	The obvious explanation for this appearance would be the b/p bulge, however the X-shape extends much further, well into the disc.
	To investigate this further, we show in Figure \ref{fig_met_bp_profiles} [Z/H] profiles along the major axis of the bar at different vertical heights. Only particles within $|y|<1$ kpc are included ($y$ is along the bar intermediate axis).
	The profile in the midplane of the model ($|z|<0.33$~kpc) is roughly flat up to ${\sim}\,3$ kpc and rises further out.
	The second [Fe/H] profile shows the b/p region ($0.33<|z|<1.0$ kpc), 
	growing from the centre outwards, reaching its maximum at $x \approx 2.5$~kpc, and then declining along $x$ further out.
	This particular behaviour coincides with the extent of the b/p bulge, (see the density contours in the middle panel of Fig. \ref{fig_met_maps} and also \citealt{blana2018}, Fig. 19), which has a length of about $2.5$--$3$ kpc and a height of about $1$ kpc.
	The third profile corresponds to the part of the model above the b/p bulge ($1.0<|z|<1.67$~kpc).
	In this slice [Fe/H] grows from the centre outwards, reaching the largest values at $x\approx 4.5$ kpc.
	All three profiles combined show that the vertical [Z/H] profile at $x\approx 2$--$3$ kpc (the end of the b/p bulge) is effectively flat up to $z\approx 1$~kpc and declines at larger heights.
	This demonstrates that the b/p bulge significantly affects the metallicity distribution in the centre of M31.
	
	Clearly, the flaring of the [Z/H] distribution further out is unrelated to the b/p bulge.
	It is possible that the metal-rich disc of M31 was thickened due to the recent merger inferred from the age-velocity dispersion relation of the disc \citep{bhattacharya2019}.
	
	In the end-on view (the right panel of Fig. \ref{fig_met_maps}), the most prominent feature is the asymmetry with respect to the $y=0$ kpc line, which we discussed above.
	
	\begin{figure}
		\centering
		\includegraphics{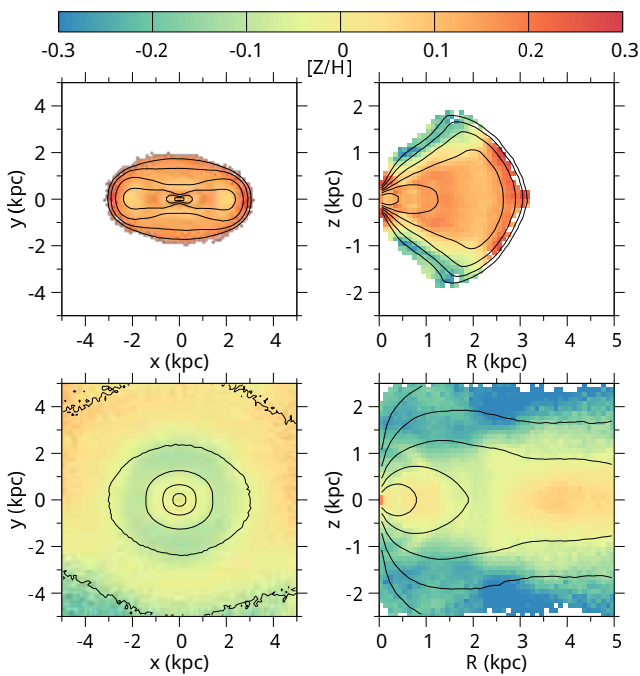}
		\caption{
			Metallicity distribution of the bar-following orbits (top row) and the non-bar-following orbits (bottom row) in the face-on view (left column) and in the meridional plane (right column).
			Overplotted are the surface density contours, in ranges $\log(\Sigma/(\mathrm{M}_\odot\,\mathrm{kpc}^{-2}))=7$--$9.5$ (top row) and $8$--$10$ (bottom row), in both cases with a log-step of $0.5$.
		}
		\label{fig_met_orbits_split}
	\end{figure}
	
	To further investigate the relation of [Z/H] and the bar, we split the stellar particles into two groups, using a very simple criterion.
	As bar-following we considered all the particles on elongated orbits, defined as $|y|_\mathrm{max} < 0.7|x|_\mathrm{max}$ and $|x|_\mathrm{max}<4$ kpc, where $|\cdot|_\mathrm{max}$ denotes the maximum absolute excursion along either major or intermediate axis of the bar (in the rotating bar frame).
	The first condition ensures that the orbits are elongated rather than circular, and the second that we only consider the particles that do not leave the bar area.
	In the second, non-bar-following group we included all the other stellar particles, i.e. those on more circular orbits in the bar region as well as those in the outer disc.
	Here we do not distinguish between the disc particles and the classical bulge particles.
	In Figure \ref{fig_met_orbits_split} we show the metallicity distribution of the two groups in the face-on view (in the bar frame) and in the meridional plane $(R,z)$. 
	We show only the pixels where a given component is present and the uncertainty is not too high, i.e. $\sigma_\mathrm{[Z/H]}<0.2$ dex.
	The latter concerns mostly the bar-following component close to the $z$-axis, due to its low density in that area.
	Similarly to Fig. \ref{fig_met_maps}, the maps were time-smoothed with $\tau=0.19$ Gyr.
	
	The bar-following orbits are significantly more metal-rich than the second group, supporting our conclusion of the bar being more Z-enhanced.
	The bar ends stand out as especially metal rich, where [Z/H] reach $\sim0.2$--$0.25$ dex, and we believe this is a robust prediction of our model.
	On the $(R,z)$ plane the boxy/peanut bulge causes a vertical flaring of metallicity.
	
	The non-bar-following population also flares, but beyond the bar end more data with extended spatial coverage is needed to establish this more firmly.
	The previously discussed metallicity-deserts perpendicular to the bar appear to be due to more metal-poor stars on nearly axisymmetric orbits in a radial range of about $R\sim1.5$--$2$ kpc.
	In the innermost $1$ kpc one can notice a [Z/H]-enhancement of a roughly donut-like shape, which could be caused by additional star formation and enrichment in a nuclear ring.
	
	\subsection{Enhancement of [\texorpdfstring{$\alpha$}{α}/Fe] in M31}
	\label{sec_alpha}
	
	\begin{figure}
		\centering
		\includegraphics{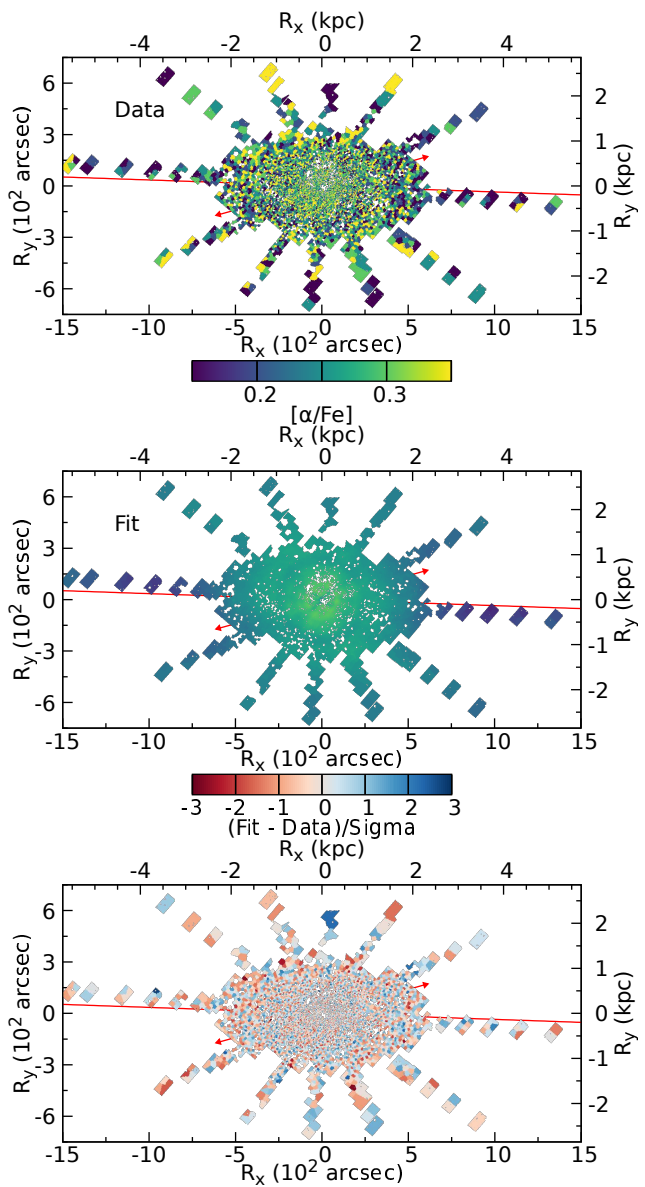}
		\caption{Map of [$\alpha$/Fe] enhancement in the Andromeda Galaxy.
			Top: the measured [$\alpha$/Fe].
			Middle: the best model.
			Bottom: standardised residuals.
			The solid red line marks the position angle of the projected disc major axis, while the red arrows mark the orientation and the extent of the bar.}
		\label{fig_alpha_obs}
	\end{figure}
	
	We now focus on the [$\alpha$/Fe] enhancement, which provides a signature of star-formation timescales \citep[e.g.][]{tinsley79, matteucci_greggio86, ferreras_silk2002,thomas2005}, i.e. larger $\alpha$-enhancement signals quicker star formation. 
	We construct separate models for the [$\alpha$/Fe] distribution, i.e. the model of the $\alpha$-enhancement is not influenced by the metallicity model, they only share the common underlying dynamical model.
	As before for metallicity, we plot in Figure \ref{fig_alpha_obs} the data from \citet{saglia2018}, our fitted best model, and a map of standardised residuals.
	Overall, $\chi^2_\mathrm{red} = 0.91$, indicating a good fit.
	There are no correlations apparent in the distributions of positive and negative residuals.
	The possible values of the particle [$\alpha$/Fe] in the model are limited to the range from $-0.3$ to $0.5$ dex, as in \citet{saglia2018}.
	As before, we tried a range of initial vertical profiles (see Eq. \ref{eq_init}).
	The preferred values of $G$, $N$ and the respective  $1\sigma$ ranges were determined similarly as for the mock data, resulting in $G=0.16\pm0.41$ dex/kpc and $N=0.42\pm0.20$ dex and a corresponding error range for the final model.
	
	At first glance we can see from these maps that the centre of M31 is relatively more $\alpha$-enriched and the shape of this feature is elongated along the projected minor axis of the galaxy.
	This is because the lower-$\alpha$ bulge material is elongated along $R_x$ \citep[see][]{saglia2018}.
	Furthermore, [$\alpha$/Fe] seems to decrease along the projected major axis for $|R_x|>2$ kpc, but is quite flat in the $R_y$ direction.
	
	\begin{figure}
		\centering
		\includegraphics{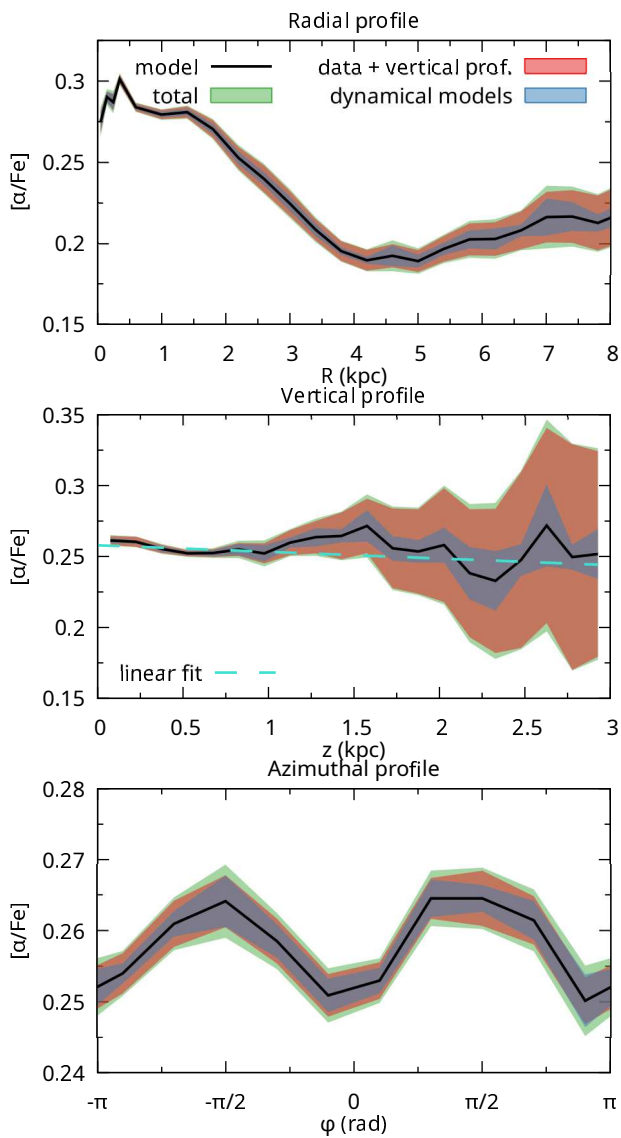}
		\caption{Profiles of the $\alpha$-enhancement, calculated from the fitted model.
			From top to bottom: the radial profile (as a function of the cylindrical radius), the vertical profile and the azimuthal profile.
			Coloured bands show the different uncertainties as in Fig.~\ref{fig_met_profiles}.
			In the middle panel the turquoise dashed line represents an uncertainty-weighted linear fit to the model curve.}
		\label{fig_alpha_profiles}
	\end{figure}
	
	Profiles of [$\alpha$/Fe], calculated from our best M2M model, are shown in Figure \ref{fig_alpha_profiles}, as a function of the cylindrical coordinates $R$, $z$ and $\varphi$.
	Here we took into account only particles within \mbox{$|z|<3$~kpc}.
	For the vertical profile we integrated over the region $R<5$ kpc, while for the azimuthal profile we considered only the region $R<5$ kpc, $|z|<3$~kpc.
	As in the case of [Z/H] profiles, we plot different types of the uncertainty bands. 
	In red we show the sum of the errors due to the data uncertainties and the range of initial vertical profiles.
	In blue we depict the uncertainty due to the underlying dynamical models, which is subdominant with respect to the other two, and
	in green we show the the overall uncertainty of the model profiles.
	
	In the inner $1.5$ kpc the radial profile is rather flat, then it decreases to a minimum at $R\ {\approx}\ 4$ kpc. Further out, the profile gently bends upwards, however, still consistent with a constant value. Recall that the data with good spatial coverage extend to deprojected $R\ {\sim}\ 5.5$ kpc, along the minor axis.
	The drop in the innermost ${\sim}\,300$ pc can be attributed to the presence of a~young stellar population \citep{saglia2018}.
	It is worth noticing that the data are consistent with M31's inner disc having super-solar $\alpha$-enhancement, and, hence, short star formation timescale \citep{thomas2005}.
	Conversely to the metallicity, in the azimuthal profile $\alpha$ is higher in the direction perpendicular to the bar, suggesting that some stars in the bar formed later than the stars in the inner disk.
	
	We made a linear fit to the vertical [$\alpha$/Fe] profile and obtained a value of $\nabla_z [\alpha\mathrm{/Fe}] =(-0.005\pm0.003) $ dex/kpc.
	Judging from the error band, [$\alpha$/Fe] is consistent with constant in the direction perpendicular to the disc, at least on average in the inner 5 kpc.
	
	\begin{figure*}
		\centering
		\includegraphics{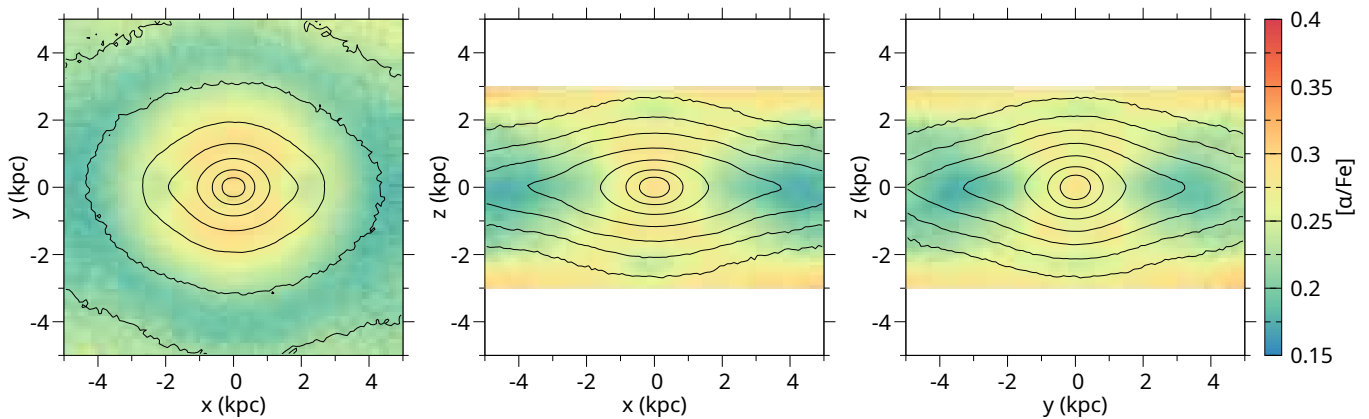}
		\caption{Deprojected maps of the mass-weighted mean [$\alpha$/Fe] in M31. From left to right: face-on, edge-on and end-on view. 
		Overplotted are the surface density contours in a range $\log(\Sigma/(\mathrm{M}_\odot\,\mathrm{kpc}^{-2}))=8$--$10$ with a multiplicative step of $10^{1/3}$.}
		\label{fig_alpha_maps}
	\end{figure*}
	
	In Figure \ref{fig_alpha_maps} we show \emph{deprojected} maps of the mass-weighted mean $\alpha$-enhancement.
	Similarly to Figure \ref{fig_met_maps}, the bar major axis is aligned with the $x$-axis and the disc rotation axis coincides with the $z$-axis.
	As before, the maps were time-smoothed following Eq. \eqref{eq_smth}.
	
	In the face-on view, the orientation of the [$\alpha$/Fe] enrichment is clearly perpendicular to the bar direction and coincides with the metallicity deserts discussed in the previous subsection.
	The face-on map exhibits a lower-$\alpha$ ring at $R\sim3-4$~kpc, signalling additional late enrichment, which is coincident with the metallicity enhanced ring visible in the bottom part of the left panel of Fig. \ref{fig_met_maps}.
	
	In the central parts of the edge-on and the end-on view the enhancement is high, with small-scale patterns which may or may not be real.
	The high [$\alpha$/Fe] are most probably related to the classical bulge and the regions where it dominates over the disc population.
	Further out, the increase to higher [$\alpha$/Fe] in the X-shaped pattern in the edge-on view is spatially approximately coincident with the decrease of metallicity outside the b/p bulge; see Fig. \ref{fig_met_maps}.
	
	In the outer part of the disc, the $\alpha$-enhancement is smaller closer to the disc plane and grows with height.
	This trend can be interpreted as an indication of the presence of an $\alpha$-rich thick disc.

	\begin{figure}
		\centering
		\includegraphics{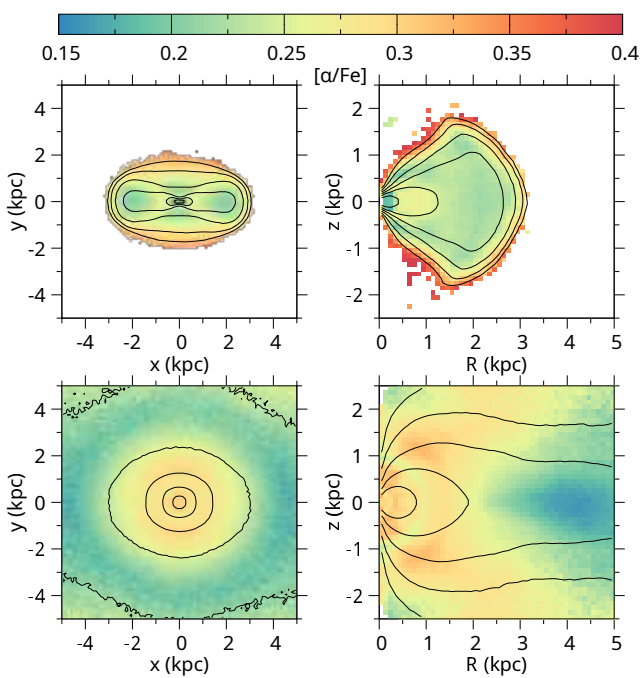}
		\caption{
				Distribution of [$\alpha$/Fe] enhancement of the bar-following orbits (top row) and the non-bar-following orbits (bottom row) in the face-on view (left column) and in the meridional plane (right column).
				Overplotted are the surface density contours, in ranges $\log(\Sigma/(\mathrm{M}_\odot\,\mathrm{kpc}^{-2}))=7$--$9.5$ (top row) and $8$--$10$ (bottom row), in both cases with a log-step of $0.5$.
		}
		\label{fig_alpha_orbits_split}
	\end{figure}
	
	Similarly to the metallicity, we also traced the $\alpha$\nobreakdash-en\-hance\-ment distribution of the bar-following ($|y|_\mathrm{max}<0.7|x|_\mathrm{max}$ and $|x|_\mathrm{max}<4$ kpc) and the non-bar-following (remaining) stellar particles.
	In Figure \ref{fig_alpha_orbits_split} we show their [$\alpha$/Fe] maps in the face-on projection and in the meridional plane $(R,z)$.
	We plot only the pixels where a given component is present and where the uncertainty is not too high, i.e. $\sigma_{\mathrm{[}\alpha\mathrm{/Fe]}}<0.15$ dex.
	The latter concerns mostly the bar-following component close to the $z$-axis.
	As in the previous instances, the maps were time-smoothed with $\tau=0.19$ Gyr.
	
	In the face-on map, the most elongated orbits, close to the major axis of the bar, have somewhat lower [$\alpha$/Fe], which could be linked to younger ages \citep[see simulations analysed in][]{fragkoudi2020}.
	This would also influence the b/p bulge, and correspondingly the bar-following orbits in the meridional plane appear to have slightly lower $\alpha$-enhancement than the non-bar orbits in the inner 2 kpc, by $\sim 0.1$ dex.
	We consider the increase of [$\alpha$/Fe] in the outer envelope of the bar-following population possible, but one must keep in mind that the uncertainty in this region is higher, chiefly due to the lower density of this component in this region.
	
	In the non-bar population one can see the low-$\alpha$ ring at $R\sim4$~kpc, which has also elevated [Z/H].
	The $\alpha$-rich region at $R\sim 1.5$ kpc is related to the metallicity desert.
	Further in, at $R\sim 1$ kpc, the model has a possible quasi-spherical shell with relatively lower $\alpha$.
	It is unclear what the significance of this feature is, but it corresponds to a higher-metallicity feature in the same region and there is a possible (noisy) lower-$\alpha$ circular feature in the data at $R\sim250$ arcsec (top panel of Fig. \ref{fig_alpha_obs}) that could cause it.
	If this shell feature is real, the most plausible explanation would be due to accreted material, but this needs further investigation.
	
	\section{Discussion}
	\label{sec_discussion}
	
	\subsection{Comparison to \texorpdfstring{\citet{saglia2018}}{Saglia et al. (2018)}}
	
	Several features of the stellar population distribution in M31 were
	clear from inspection of the maps obtained by \citet{saglia2018},
	such as the [Z/H] enhancement along the bar and the [$\alpha$/Fe] peak in the
	central 1 kpc.
	In addition, \citet{saglia2018} constructed a simple decomposition of their stellar population data into various components (classical bulge, b/p bulge, bar and disk), which was based on using different regions of the sky to constrain
	the classical bulge and bar components in the mass model of \citet{blana2018}.
	
	It is instructive to compare our detailed results to the mean profiles of their model (blue and black lines in Figs. 21--24 of \citealt{saglia2018}).
	We found a steep increase of metallicity and steep decrease of [$\alpha$/Fe] in the inner $200$ pc ($50\arcsec$), which was directly mandated by the data we used.
	\citet{saglia2018} ascribed the [Z/H] peak to the classical bulge, while the $\alpha$ decrease was attributed to the recent star formation.
	We agree with the \citet{saglia2018} model that along the bar metallicity is constant, while [$\alpha$/Fe] decreases.
	They found almost flat metallicity profiles for the b/p bulge and the disc (Figs. 23--24).
	Our model showed that the b/p bulge has an X-shape in [Z/H], as one can find in Fig. \ref{fig_met_maps}.
	Thus, the metallicity profiles in the b/p bulge grow with radius, up to its end at $\approx 2.5$ kpc, at a range of heights $z\sim0.3$--$1$ kpc.
	The [Z/H] morphology of the disc shows diverse features: metallicity deserts perpendicular to the bar at $|y|\sim2$~kpc, then further away an enriched elongated ring at $R\sim3$--$5$~kpc and a possible further decrease beyond $5$ kpc.
	The vertical distribution of [Z/H] beyond the b/p bulge seems to be flaring.
	Our [$\alpha$/Fe] distribution in the central part of M31 show enhancement perpendicular to the bar and up to a significant height.
	Our model show a steeper $\alpha$-overabundance decrease up to $R\sim4$~kpc than in Fig.\ 24 of \citet{saglia2018}.
	
	It would be tempting to perform here a decomposition into classical bulge, b/p bulge, bar and disk components,  using the model by \citet{blana2018} that explicitly consists of two types of particles (belonging to the classical bulge and disc).
	However, because of the spin-up through angular momentum transfer by the bar \citep{saha2012, saha2016}, some of the classical bulge orbits overlap with those of the bar.
	Our current method cannot differentiate between the mean metallicities of the two components on the same orbit.
	From our final model, we can only tentatively say that the bulge component has a very steep gradient, while the profile of the disc component (which contains the bar) is flat or even decreasing towards the centre, however we are not able to give uncertainties.
	The classification of the stellar particles as bar-following and non-bar-following, presented in Figs. \ref{fig_met_orbits_split} and \ref{fig_alpha_orbits_split}, should not be taken as a proxy for a decomposition into the classical bulge and the combined bar and b/p bulge, due to the aforementioned spin-up of the classical bulge and overlap of orbits.
	Improving this aspect of the model seems possible but is beyond the scope of this paper.
	
	Compared to the earlier modelling, several new qualitative results have emerged from our full particle modelling of the stellar population data: the discovery of a metallicity X-shape in the edge on-view, caused by metallicity enhancement in the b/p bulge and the further flaring in the disc. Additionally, we found a [Z/H]-enhanced, low-$\alpha$ ring, as well as metallicity maxima at the tips of the bar.
	Furthermore, we found indications of an $\alpha$-enhanced thick disc in M31.
	
	\subsection{Comparison with other galaxies}
	
	The impact of the bar on the stellar population gradients has been an active topic for some time now.
	It has been found that the profiles along the bars are flatter than those perpendicular to bars \citep{sanchez-blazquez2011, fraser-mckelvie2019, neumann2020}.
	Interestingly, the results of \citet[Fig. 2]{fraser-mckelvie2019} indicate that the gradients may be either positive or negative \citep[see also][]{perez2009}, but the signs of the gradients in the directions parallel and perpendicular to the bar remain correlated with each other.
	However, judging from our M31 results, it is important to go beyond simple linear fitting, since the profiles in the bar region can be more complicated.
	Looking at Fig. \ref{fig_met_profiles}, one would infer a negative radial [Z/H] gradient in the central $\sim2$ kpc and a positive one further out, up to the bar end.
	\citet{seidel2016} in their Fig. 10 presented a step in this direction, where they found that the slopes of the Mg$b$ index profiles change abruptly at around 10--15\% of their bar length.
	
	It appears that the metallicity enhancement close to the bar ends, as we find in our M31 model, has not yet been widely appreciated.
	However, a closer inspection of available spatially resolved data sometimes reveals such trends.
	Many of the BaLROG bars show this feature \citep{seidel2016}, as well as some of the galaxies analysed in the TIMER project \citep{gadotti2019, neumann2020}.
	Moreover, the APOGEE data for the Milky way indicate the presence of a metallicity maximum close to the bar end at $l\approx 30\degr$ \citep[e.g.][]{ness_freeman2016}. The subsequent maps of \citet{bovy2019} can be interpreted as that [Fe/H] is either enhanced at the bar end or in a nearby ring.
	On the other hand, \citet{wegg2019} found more enhanced metallicities in their bar fields 4 kpc from the centre.
	
	It is also interesting to compare side-on views of our results to other published examples.
	According to the M31 models, [Z/H] is enhanced in the b/p bulge and shows signs of flaring further away from the centre, while [$\alpha$/Fe] is high in the classical bulge region and at large heights above the disc plane.
	The edge-on galaxy NGC 1381 of \citet{pinna2019a} shows clear signs of boxy/peanut shape \emph{both} in metallicity and [Mg/Fe].
	\citet{williams2011} suggested that this galaxy may have a small classical bulge, which would make it different from M31 with its sizeable classical bulge.
	\citet{pinna2019b} studied another galaxy with a b/p bulge, FCC 177.
	However, this one does not have obvious signs of the b/p bulge in metallicity and certainly not in [Mg/Fe].
	However, curiously, [Fe/H] seems to have maxima in the disc plane at $\approx3$ kpc from the galaxy centre, beyond the b/p bulge (of size $\sim1.5$--$2$ kpc, judging from the isophotes).
	In case of the Milky Way the model of \citet[Fig.\ 10]{portail2017b} suggests that the [Fe/H] distribution in the centre has a peanut-like shape (see also the $N$-body models of \citealt{debattista2017} and \citealt{fragkoudi2018}).
	
	In the Milky Way $\alpha$-enhancement is anti-correlated with metallicity, broadly speaking \citep[e.g.][and many others]{hayden2015}.
	In our model this is also true to a certain degree.
	In the face-on view $\alpha$-rich regions coincide with the metallicity deserts perpendicular to the bar.
	The high-[Z/H] ring is relatively less $\alpha$-enhanced.
	In the edge-on views, at large distances from the centre, a negative metallicity gradient corresponds to a positive $\alpha$ gradient.
	Apparently, the centre of M31 does not follow this trend, presumably due to the metal-rich, high-$\alpha$ classical bulge.
	
	\subsection{Origins of the spatial trends}
	
	Many of the models for the origin of the stellar populations in disk galaxies are focused on the Milky Way, because it is the galaxy for which we have the most detailed data.
	However, these models are often useful for M31 as well, since it is a disk galaxy of similar mass and also includes a bar and b/p bulge component.
	Often, the authors assume a certain initial metallicity distribution and test the impact of the further secular evolution \citep[e.g.][]{martel2013, martinez-valpuesta_gerhard2013, dimatteo2013}.
	The other approach is tracing stellar populations in hydrodynamical simulations with star formation and feedback \citep[e.g.][]{grand2016, debattista2017, debattista2019, tissera2016,tissera2017,tissera2019,taylor_kobayashi2017}.
	
	\citet{dimatteo2013} analysed a case where a disc galaxy has initially an axisymmetric distribution of mass and metallicity,
	with [Fe/H] assigned such that it decreased linearly with the radius.
	During the further evolution a bar and spiral arms formed.
	The creation of the bar induced outwards radial migration and effectively mixed the stellar metallicities from the galaxy centre up to the bar length, creating an enhancement along the bar and metallicity-deserts perpendicular to it \citep[see also][Fig. 15]{debattista2020}.
	This resulted in an azimuthal variation of metallicity.
	They found that the ratio of the azimuthal variation $\delta_\mathrm{[Fe/H]}$ (measured e.g. in the bar region, see \citealt{dimatteo2013}, Fig.~8 and Sect. 3.3) and the initial metallicity gradient $\Delta_\mathrm{[Fe/H]}$ can be approximated by $\delta_\mathrm{[Fe/H]}/\Delta_\mathrm{[Fe/H]}{\sim}1\pm 0.5$, independently of the initial gradient.
	In our case $\delta_\mathrm{[Z/H]}\approx 0.02$ dex (Fig.~\ref{fig_met_profiles}),
	which would correspond to an initial radial gradient in Andromeda of the order of $\nabla_R [Z/H]{\sim}0.02\pm0.01$ dex/kpc. This is compatible with the metallicity gradient in the outer disc of M31 derived from the PHAT survey \citep{gregersen2015}, but also with typical radial gradients in other galaxies \citep{sanchez-blazquez2014, goddard2017, zheng2017manga} and in some hydrodynamical simulations for redshift $z<1$ \citep{tissera2017}.
	Note that the exact shape of the bar-related enhancement depends on the initial mix of the stellar populations \citep{khoperskov2018variation}.
	
	Further work is needed to understand the influence of the purported recent merger on M31 \citep{hammer2018, bhattacharya2019}, and whether it deposited enough material in the bar region to significantly change the mean stellar population parameters there.
	
	Alternatively, the enhancement along the bar may be caused by the so-called kinematic fractionation effect \citep{debattista2017,fragkoudi2017a}, by which the population that is colder before the bar formation would become more strongly aligned with the bar afterwards.
	If the colder population was additionally more metal-rich, this would result in a metallicity enhanced bar.
	In the Milky Way the more metal-rich populations exhibit distributions that are more strongly barred \citep{portail2017b}.
	Moreover, the bar metallicity can be further enhanced during its growth, if the bar can capture new stars that are more metal rich \citep[e.g.][]{aumer_schonrich2015}.

	For the $\alpha$-enhancement in M31 we find only a weak anticorrelation with the bar.
	The stellar component on the most elongated bar-following orbits is somewhat more $\alpha$-enhanced than the surrounding disc.
	The high $\alpha$ value ($\gtrsim 0.15$ dex everywhere) implies that the stars in the entire bar region must have formed quite rapidly at early times \citep{pipino2006, pipino2008}.
	The stellar population that formed the bar also appears to not have had an initial [$\alpha$/Fe] radial gradient, i.e.\ the entire bar and the surrounding disk stars must have formed rapidly at early times.
	The original galaxy also should not have had multiple discs of different $\alpha$, such as a distinct $\alpha$-rich thick disc, since this would probably also result in a bar-related non-axisymmetry \citep{fragkoudi2018}.
	The slightly lower [$\alpha$/Fe] in the bar region could be related to late-time star formation, which is also necessary for the formation of the $\alpha$-poor ring.
	
	The X-shaped side-on view of the metallicity is an expected outcome of the hydrodynamical simulations, both in isolation \citep{debattista2017} and in the cosmological context \citep{debattista2019, fragkoudi2020}.
	The X-shape in the region of our model's b/p bulge ($|x|\lesssim 2.5$ kpc) is probably somewhat weaker than in the simulated galaxies, since the dynamical b/p bulge is weaker due to the presence of the classical bulge.
	Different from the models, the metallicity distribution in M31 keeps flaring with growing distance from the centre.
	The origin of this flare could be related to the heating of the disk caused by the merger inferred to have occurred $\sim3$~Gyr ago \citep[][]{hammer2018, bhattacharya2019}.
	On the other hand, the cosmological zoom-in model of \citet{rahimi2014} also exhibits a positive radial metallicity gradient at large heights. 
	The authors ascribed it to the flaring of the young, metal-rich stellar population in the outer disc.
	Certainly, more spatially extended data beyond the bar end would help to constrain the model more strongly.
	The side on-map of [$\alpha$/Fe] in M31 looks different from those obtained in the Auriga simulations \citep{fragkoudi2020}.
	Note that these Auriga galaxies do not have classical bulges (as inferred from small S\'ersic indices), while M31 clearly has one. Note also that their average [$\alpha$/Fe] are much closer to the solar value than in Andromeda.
	
	Metallicity enhancement along bars and maxima at their ends are also found in hydrodynamical simulations \citep{debattista2019, fragkoudi2020}.
	In particular, two of the runs (Au18 and Au23) analysed by \citet{fragkoudi2020} exhibit both [Fe/H] maxima at the bar ends and rings of enhanced metallicity and relatively lower-$\alpha$.
	The authors of that study relate the rings to on-going star formation in the region between the bar end and the corotation radius.
	The runs Au18 and Au23 are different with respect to the others analysed by \citet{fragkoudi2020} also in terms of their rotation curves, which are not centrally peaked, similarly to M31 \citep{blana2018} and the Milky Way \citep[e.g.][]{portail2017a}. The Auriga bars are also significantly smaller than their corotation radii, i.e., they are relatively slower.
	The enhancement close to the tip of the bar could be related to the star formation activity observed there in some galaxies \citep[e.g.][]{phillips1996}. On the other hand, 
	\citet[Fig. 15]{debattista2020} found that if one tags the particles with metallicity corresponding to an axisymmetric galaxy and let the galaxy form a bar, then the metallicity becomes enhanced along it, but the maxima at the bar ends do not form.
	
	High metallicity rings and arcs beyond the metallicity desert have also been found in \citet{khoperskov2018variation}, where they are related to spiral arms \citep[see also][Fig. 8]{debattista2019}.
	They arise from stellar populations with different kinematics and metallicity patterns, which respond differently to the spiral arms.
	The ring we find in our model in principle could have a similar origin.
	
	\section{Conclusions}
	\label{sec_conclusions}
	
	In this paper we devised a new chemodynamical technique based on the made-to-measure (M2M) modelling framework, enabling us to constrain the distribution of galactic stellar populations.
	As input the method uses mass-weighted maps of a mean stellar population parameter, e.g.\ metallicity, age, or $\alpha$-enhancement.
	We start with a dynamical $N$-body model, constrained by surface density and kinematics.
	Then we tag the particles with single values of e.g.\ metallicity, corresponding to the mean metallicities along their orbits, and then we adjust those values to match the observational constraints.
	We tested our method on mock data and found that the radial and azimuthal profiles in the plane of the galaxy are well reproduced, while in the case of the vertical profile we are only able to obtain an average gradient, without finer details.
	
	We applied our technique to the Andromeda Galaxy.
	We used [Z/H] and [$\alpha$/Fe] maps derived by \citet{saglia2018} to constrain the three-dimensional distribution of metallicity and $\alpha$-enrichment in the context of the M31 dynamical model of \citet{blana2018}.
	
	Our main results can be summarised as follows:
	\begin{enumerate}
		\item We find that the metallicity is enhanced along the bar, while perpendicular to the bar we see a clear depression, akin to the so-called star formation deserts \citep{james2009, donohoe-keyes2019}.
		The enhancement along the bar is directly related to the high metallicities of bar-following orbits.
		We also find that the metallicity distribution exhibits a peak at the radius corresponding to the bar size, corresponding in the face-on view to an elongated, [Z/H] enhanced ring.
		
		\item Closer inspection of the side-one view and orbital analysis reveals that the [Z/H] enhancement has an X-shape caused by the boxy/peanut bulge.
		The average vertical [Z/H] gradient in the entire inner region of M31 is $-0.133\pm0.006$ dex/kpc. This is significantly lower (by about half) than 
		typical values in other galaxies presented by \citet{molaeinezhad2017} but
		there are some galaxies with similar vertical gradients in their sample.
		
		\item On average, the [$\alpha$/Fe] enhancement in Andromeda is high compared to solar, indicating short formation timescales.
		The centre of M31 is more $\alpha$-enhanced than the surrounding disc, likely due to the presence of the classical bulge.
		We find a relatively lower-$\alpha$ ring, corresponding to the metallicity ring.
		Some of the most elongated bar-following orbits also have a somewhat lower [$\alpha$/Fe] than the average.
		The average vertical [$\alpha$/Fe] gradient is flat, however a closer look reveals more structure: 
		near the centre $\alpha$-enhancement decreases slightly while in the disc it increases with height; this may signify a presence of a more strongly $\alpha$-enhanced thick disc.
	\end{enumerate}

	From the data of \citet{saglia2018} and our model the following formation pathway for the Andromeda Galaxy can be deduced.
	Given the overall high level of $\alpha$-enhancement, the stars in both the bulge region and the inner disc must have formed relatively quickly, with faster star-formation in the bulge region.
	Since the current structure of the metallicity and $\alpha$ distributions are qualitatively different, they must have been different initially.
	The distribution of [$\alpha$/Fe] is consistent with no radial gradient or any other structure in the original disc.
	The galaxy assembly resulted in a sharp peak of metallicity in the central few hundred parsecs and a more gentle negative gradient in the remaining disc.
	The formation of the bar lead to a re-arrangement of the [Z/H] distribution, causing the formation of metallicity-deserts and a flat gradient along the bar.
	Afterwards, the star formation continued close to the bar ends in the leading edges of the bar, producing metallicity enhancements in the ansae and the [Z/H] enhanced, lower-$\alpha$ ring.
	The star formation in the very centre at $\sim200$ pc also continued, however it was rather mild, leading to only a small decrease of [$\alpha$/Fe] there.
	Andromeda probably experienced recently a fairly massive minor merger \citep{hammer2018, bhattacharya2019}, which most probably lead to flattening of the gradients \citep[e.g.][]{taylor_kobayashi2017}.
	It would be interesting to further investigate the impact of a merger on the distribution of the stellar populations in barred galaxies.
	
	In this work we constrained our models with spatially resolved maps of line-of-sight averages of a given stellar population parameter.
	Thus we could only constrain the mean metallicity and [$\alpha$/Fe] distribution in the corresponding parts of the orbit-space and deprojected space.
	The obvious way forward is to use separate maps for intervals of metallicity or age from full spectral fitting, similar to those presented in \citet[Fig. 1]{peterken2020}.
	
	Our newly developed technique proved to be a valuable tool for studying the distribution of the stellar populations in galaxies, 
	To date, analysis was always confined to looking at galaxies either face-on or edge-on.
	Here we are able to model the three-dimensional distribution and thereby connect the results from both perspectives.
	In the future, we plan to apply our technique to other galaxies with spatially-resolved maps of metallicity, $\alpha$-enhancement and age, thereby obtaining a more complete picture of the chemodynamical structures in disk galaxies.
	
	\begin{acknowledgements}
		We would like to thank Chiara Spiniello for useful discussions about stellar population measurements and the anonymous referee for constructive comments.
		The research presented here was supported by the
		Deutsche Forschungsgemeinschaft under grant GZ GE 567/5-1,
		by the National Key R\&D Program
		of China under grant No. 2018YFA0404501, and by the National Natural Science
		Foundation of China under grant Nos. 11761131016, 11773052, 11333003.
		MB acknowledges CONICYT fellowship Postdoctorado en el Extranjero 2018 folio 74190011 No 8772/2018 and the Excellence Cluster ORIGINS funded by the Deutsche Forschungsgemeinschaft (DFG, German Research Foundation) under Germany's Excellence Strategy - EXC-2094-390783311.
		LZ acknowledges the support of National Natural Science Foundation of China under grant No. Y945271001.
	\end{acknowledgements}
	
	\bibliographystyle{aa.bst}
	\balance 
	\bibliography{bibliography}
	
	\begin{appendix}
		
		\section{Impact of the vertical prior at low inclination}
		\label{app_vertical_prior}
		
		\begin{figure}
			\centering
			\includegraphics{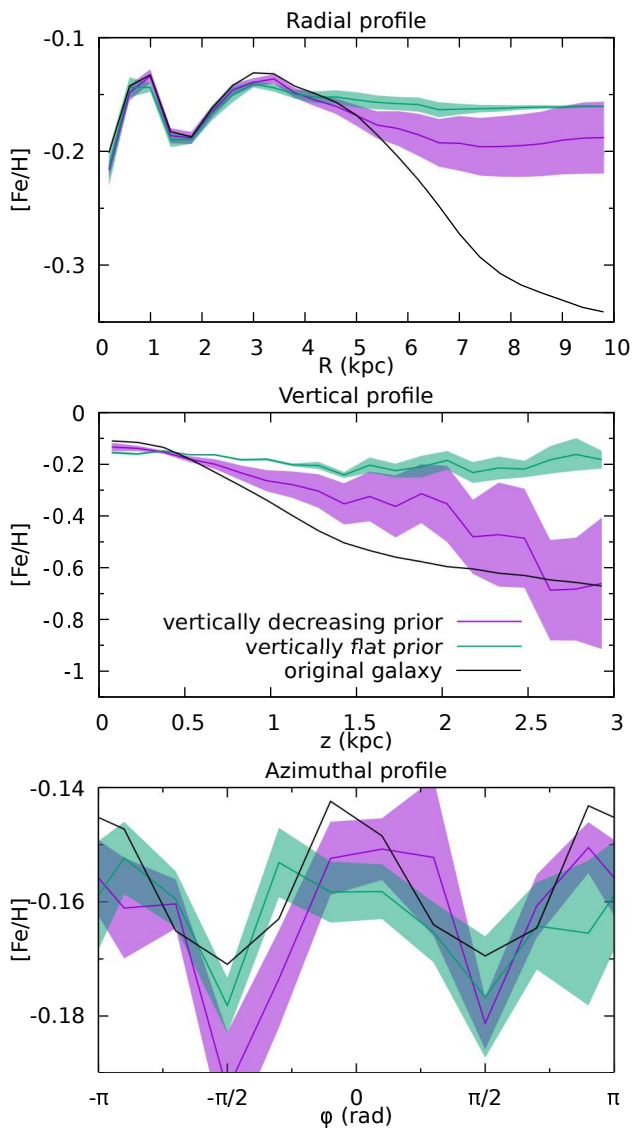}
			\caption{
				Profiles of [Fe/H] for the mock test for an inclination of $45\degr$.
				From top to bottom: radial profiles (as a function of the cylindrical radius), vertical profiles, and azimuthal profiles.
				The green lines depict effects of the modelling assuming a flat vertical prior. 
				The violet lines correspond to the optimal, vertically decreasing prior. Coloured bands show the respective uncertainties.
				The black lines shows the profiles of the target galaxy.
				Note that the data constrains extend only to $R<6$ kpc.
			}
			\label{fig_mock_incl45_profiles}
		\end{figure}
		
		To illustrate the importance of the particle initialisation according to a vertical prior (following Eq.~\ref{eq_init}) we performed the following test. 
		Using the same procedure as in Section \ref{sec_mock_test}, we observed the model of \citet{portail2017b} not at the inclination of $77\degr$, but instead at $45\degr$, and modelled the mock data in two different ways.
		In the first approach, we initialised the metallicity of all the particles to a single value of $-0.159$ dex, which is equal to the mean over all the observed lines of sight.
		In the second approach, we followed the prescription outlined in the main part of the paper.
		Namely, we sampled a range of $G$s and $N$s and found that the model with the lowest $\chi^2$ is obtained for $G=-0.35\pm0.20$ dex/kpc and $N=-0.216\pm0.019$ dex.
		
		In Figure \ref{fig_mock_incl45_profiles} we show the results of our test.
		The green lines depict the model initialised with a constant value of the metallicity.
		The violet lines shows the model with the optimal choice of $G$ and $N$.
		Both can be compared to the black lines, which correspond to the original model of \citet{portail2017b}.
		When initialised to a constant value of metallicity, the model significantly underestimates the vertical gradient -- in fact, [Fe/H] is almost constant as a function of $z$.
		However, initialisation with a~vertical gradient brings the final vertical profile much closer to the profile of the input model.
		The uncertainty bands gauge the mismatch more faithfully too.
		Additionally, also the radial profile follows the black line more closely around $R\approx5$ kpc.
		One should notice that the radial profile is wrong beyond $\gtrsim6$ kpc because the data at the $45\degr$ inclination does not cover that part of the galaxy.
		To summarise, the initialisation of the vertical profile is an important part of our approach and it works reasonably well.
		
		\section{Impact of the radial prior}
		\label{app_radial_prior}
\begin{figure}[b]
			\centering
			\includegraphics{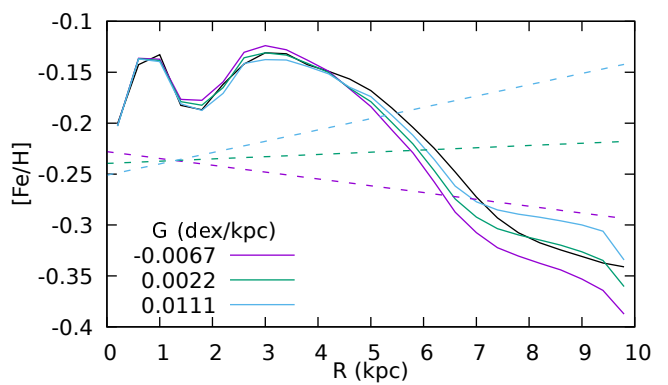}
			\caption{
				Initial (dashed lines) and final (solid lines) radial metallicity profiles of the mock galaxy.
				The green line marks the model with the lowest final $\chi^2$, while the blue and violet lines correspond to the 1$\sigma$-worse models.
				The black line shows the radial profile of the original mock galaxy.
			}
			\label{fig_mock_radial}
		\end{figure}

		In our fiducial modelling approach the initialisation of the population label depends only on the vertical coordinate.
		One could imagine imposing also a radial dependence.
		To test this possibility, we performed the same mock test as in Section \ref{sec_mock_test}, but now we impose the following initial profile for the metallicity [Fe/H]$_i$ of each particle
		\begin{equation}
			\mathrm{[Fe/H]_i}(R_i) = G(R_i-R_0)+N,
			\label{eq_radial}
		\end{equation}
		where $G$ and $N$ denote the gradient and its normalisation.
		$R_i$ is the cylindrical radius coordinate of the $i$-th particle and $R_0$ is a normalisation constant, chosen as $1.3$ kpc (to reduce correlation between $G$ and $N$). 
		The values that result in the lowest $\chi^2$ are $G=0.005\pm0.012$ dex/kpc and $N=-0.24\pm0.05$ dex.
		The preferred range of the radial gradient is tightly constrained to the null value.
		In Figure \ref{fig_mock_radial} we present the impact of the radial initialisation on the final radial profile.
		The difference is rather small, well within the uncertainties of Fig. \ref{fig_mock_profiles}.
		We also inspected the impact of the radial prior on the shape of the azimuthal profile and it turned out to be minimal.
		Thus, our approach of using the initialisation that depends only on the vertical coordinate is justified. 
		
		\section{Asymmetry of the metallicity distribution}
		\label{app_dust}
		
		There are many asymmetries in the products derived from the data collected by \citet{opitsch2018}.
		The maps of the velocity dispersion and the fourth Gauss-Hermit moment $h_4$ display asymmetry between the northern ($R_y>0$, closer to us) and the southern ($R_y<0$, farther away from us) part.
		\citet{blana2018} ascribed the asymmetries to the dust in the disc obscuring the central part of M31.
		To correct that effect, they used the \citet{draine2014} map of the dust in M31.
		Assuming that the dust is concentrated in the disc plane, they employed the \citet{draine_li2007} model to calculate the extinction along the line of sight.
		The resulting map of the fraction of the galaxy's light that reaches us is shown in the middle panel of Fig. 22 in \citet{blana2018}.
		The key feature of this dust model is that the light of the stars behind the disc plane is extincted and contributes less to the observed properties than the stars in front of the disc plane.
		
		\begin{figure}
			\centering
			\includegraphics{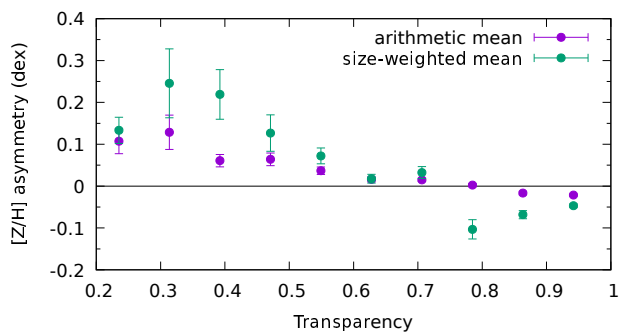}
			\caption{
				Mean metallicity asymmetry as a function of the light transparency through the dust. 
				The green points were obtained weighting each Voronoi cell by its size on the sky, while the violet points show simple arithmetic averages.
				Error bars correspond to standard errors of the mean (corrected for the effective sample size for the weighted means).
			}
			\label{fig_asymm_transp}
		\end{figure}
		
		Also the stellar populations maps of \citet{saglia2018} display north-south asymmetries, most notably H$\beta$, age and metallicity (but not $\alpha$-enhancement).
		In order to test whether the dust has an impact on the observed asymmetry of the metallicity distribution we performed the following analysis.
		For each Voronoi cell we computed the fraction of the light blocked by the dust, using the prescription outlined in the previous paragraph.
		We also calculated the [Z/H] asymmetry of each cell, i.e. for a cell located at $(R_x, R_y)$ we identified all the pixels located at $(-R_x, -R_y)$.
		The asymmetry is the difference between the cell's metallicity and its reflection.
		It is positive if the reflection has lower metallicity than the cell in question and negative otherwise.
		In Figure \ref{fig_asymm_transp} we compare the dust-related transparency and the asymmetry of the data.
		We averaged the asymmetry of the Voronoi cells in the bins of the transparency.
		Besides simple arithmetic averages, we show averages weighted by the cells' size on the sky, which accounts for significantly larger cell sizes far away from the M31 centre.
		It is apparent that at low transparency the data has positive asymmetry (i.e. higher [Z/H] than on the other side).
		
\begin{figure}[b]
			\centering
			\includegraphics{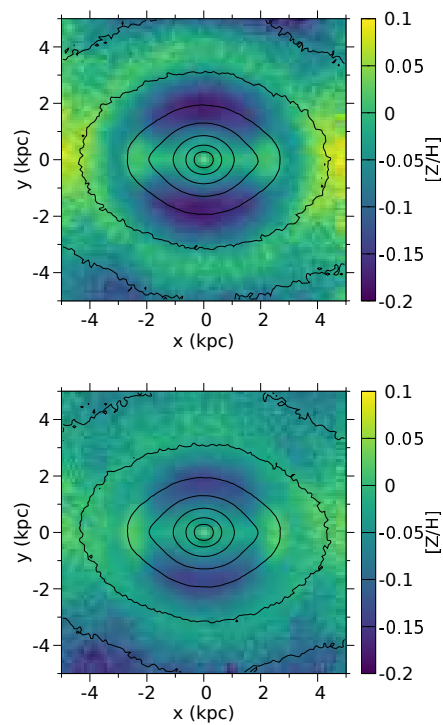}
			\caption{
				Face-on views of models fitted to two variants of the data.
				Top: southern part of M31 ($R_y<0$) point-symmetrised onto the northern part.
				Bottom: Voronoi cells fulfilling jointly two criteria (less than $80\%$ of light throughput and metallicity asymmetry bigger than $0.1$ dex) were discarded before fitting.
			}
			\label{fig_asymm_corr}
		\end{figure}

		Given that the dust is a possible source of the asymmetry, we attempted two ways of mitigating the dust impact.
		In the first approach, we assumed that the whole northern side is compromised.
		We discarded that part of the data (i.e. the cells above the red line in Fig. \ref{fig_met_maps}), while the southern part was duplicated by rotating it by $180\degr$, effectively making the data point-symmetric.
		Then, we proceed with fitting an M2M model.
		The resulting face-on map is presented in the top panel of Fig. \ref{fig_asymm_corr}.
		As expected, the map is symmetric with respect to the $y=0$ line. It displays the metallicity-enhanced ring and a [Z/H] decrease farther away from the centre.

		In the second approach we removed the Voronoi cells which fulfilled jointly two criteria: the non-attenuated light fraction was less than $80\%$ (as derived using the discussed above dust model devised by \citealt{blana2018}) and the asymmetry was larger than $0.1$ dex.
		The resulting face-on map of a model fitted to the filtered data is shown in the bottom panel of Fig. \ref{fig_asymm_corr}.
		The metallicity-enhanced ring is still clearly visible in the bottom part of the map.
		The metallicity in the top part was reduced, however is was not enough to bring it into symmetry.
		In this approach the asymmetry between the $y>0$ and $y<0$ parts was reduced from $\sim0.2$ dex to $\sim0.1$~dex.

\begin{figure}
			\centering
			\includegraphics{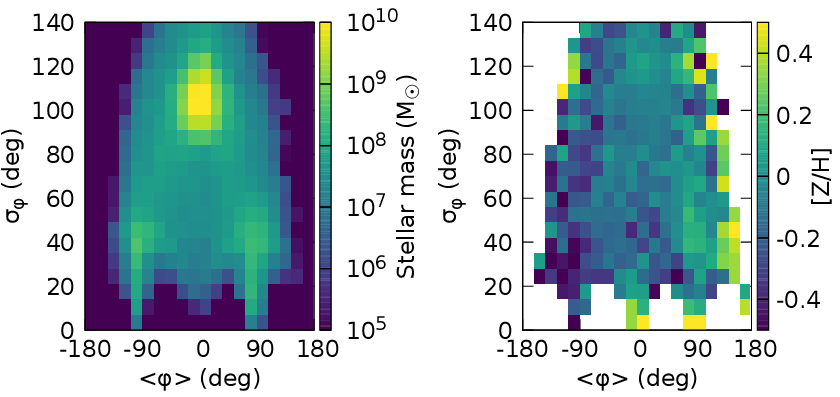}\\
			\includegraphics{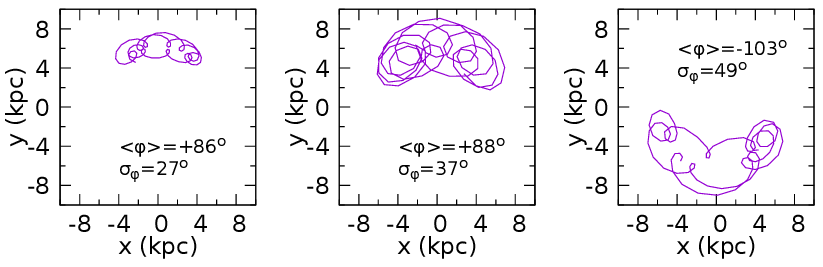}
			\caption{
				Top row: distribution of stellar orbits (left) and their mean metallicity (right) as a function of the mean position angle and its standard deviation.
				Bottom row: three examples of stellar orbits (in the rotating reference frame), revolving around the Lagrange points $L_4/L_5$.
			}
			\label{fig_asymmetric_orbits}
		\end{figure}

		As we mentioned in Section \ref{sec_met}, the asymmetric feature persists for at least $6$ Gyr when we let the model evolve freely.
		We decided to investigate which orbits are responsible for this phenomenon.
		To this effect, we evolved the model of $0.5$ Gyr and
		every $10^{-4}$ Gyr, we recorded for each particle the position angle $\varphi$ in the disc plane (in the reference frame rotating with the bar).
		Then, for each stellar orbit we calculated the mean position angle $\langle \varphi \rangle$ and its standard deviation $\sigma_\varphi$.
		We show the distribution of the stellar orbits on the plane $(\langle \varphi \rangle, \sigma_\varphi)$ and their average metallicity in the top row of Figure \ref{fig_asymmetric_orbits}.
		For an orbit symmetric with respect to the major axis of the bar we expect $\langle \varphi \rangle\approx 0\degr$.
		If the orbit is axisymmetric, then their recorded $\varphi$ are distributed uniformly, resulting in $\sigma_\varphi\approx104\degr$.
		These two numbers correspond to the highest peak in the orbital distribution.
		However, the distribution extends to much smaller values of $\sigma_\varphi$.
		In particular, the are two distinct populations of orbits with $\sigma_\varphi \lesssim 60\degr$, located at $\langle \varphi \rangle \approx \pm 90\degr$.
		The group of orbits clustered at $\langle \varphi \rangle \approx+ 90\degr$ has a higher metallicity than the group at $-90\degr$.
		We checked also the mean distance of those orbits and the result was $\approx6$~kpc, the same as the corotation radius.
		All those facts highly suggest that the orbits around $L_4/L_5$ Lagrange points are responsible for a~stable asymmetry. In the bottom row of Figure \ref{fig_asymmetric_orbits} we depicted three such orbits.
		
		Given the existence of orbits supporting the asymmetry, it is possible that the asymmetry is actually real.
		It could have arisen due to enhanced star formation in spiral arms on one side of the galaxy.
		In fact, the GALEX FUV image \citep{thilker2005} shows some asymmetry in the inner star-forming ring.
		However, if this was the case, one would also expect some asymmetry in the distribution of [$\alpha$/Fe], which is not present.
		
		To summarise, the apparent asymmetry of the M31 metallicity data, and its model, can be related to the dust obscuration in its northern part.
		We hypothesise that the bulge stars are obscured by the star-forming disc, hence the ages are biased towards the younger ones.
		This, through the well-known age-metallicity degeneracy, leads to an overestimated metallicity for the entire line-of-sight.
		The persistence of such an asymmetry in the model is supported by a presence of banana-shaped orbits related to the Lagrange points.
		On the other hand, since it is dynamically possible, it is conceivable that such a feature is actually a real one, related to e.g. ongoing asymmetric star formation activity.
		We note that some of the optically bluish galaxies presented in the appendix of \citet{seidel2016} also show asymmetries in their Lick-indices based metallicity maps.
	\end{appendix}
	
\end{document}